\documentclass[a4paper,envcountsame]{llncs2e/llncs}
\def\tikzChristmasI
{\begin{tikzpicture}[xscale=0.22,yscale=0.22]

	\draw [help lines] (0,0) grid (25,20);

    \foreach \pt in {(1,19),(3,19),(4,19),(6,19),(7,19),(8,19),(9,19),(11,19),(12,19),(14,19),(15,19),(16,19),(17,19),(19,19),(20,19),(21,19),(22,19),(24,19),(8,18),(16,18),(21,18),(1,17),(3,17),(6,17),(8,17),(9,17),(11,17),(14,17),(16,17),(17,17),(19,17),(21,17),(22,17),(24,17),(1,16),(2,16),(3,16),(4,16),(6,16),(7,16),(8,16),(9,16),(10,16),(11,16),(12,16),(14,16),(15,16),(16,16),(17,16),(18,16),(19,16),(20,16),(21,16),(22,16),(23,16),(24,16),(3,15),(8,15),(11,15),(16,15),(19,15),(21,15),(24,15),(1,14),(3,14),(4,14),(6,14),(8,14),(9,14),(11,14),(12,14),(14,14),(16,14),(17,14),(19,14),(20,14),(21,14),(22,14),(24,14),(21,13),(1,12),(3,12),(6,12),(8,12),(11,12),(14,12),(16,12),(19,12),(21,12),(22,12),(24,12),(1,11),(2,11),(3,11),(4,11),(6,11),(7,11),(8,11),(9,11),(11,11),(12,11),(14,11),(15,11),(16,11),(17,11),(19,11),(20,11),(21,11),(22,11),(23,11),(24,11),(3,10),(8,10),(16,10),(21,10),(24,10),(1,9),(3,9),(4,9),(6,9),(8,9),(9,9),(11,9),(14,9),(16,9),(17,9),(19,9),(21,9),(22,9),(24,9),(1,8),(2,8),(3,8),(4,8),(5,8),(6,8),(7,8),(8,8),(9,8),(10,8),(11,8),(12,8),(14,8),(15,8),(16,8),(17,8),(18,8),(19,8),(20,8),(21,8),(22,8),(23,8),(24,8),(3,7),(6,7),(8,7),(11,7),(16,7),(19,7),(21,7),(24,7),(1,6),(3,6),(4,6),(6,6),(7,6),(8,6),(9,6),(11,6),(12,6),(14,6),(16,6),(17,6),(19,6),(20,6),(21,6),(22,6),(24,6),(8,5),(21,5),(1,4),(3,4),(6,4),(8,4),(9,4),(11,4),(14,4),(16,4),(19,4),(21,4),(22,4),(24,4),(1,3),(2,3),(3,3),(4,3),(6,3),(7,3),(8,3),(9,3),(10,3),(11,3),(12,3),(14,3),(15,3),(16,3),(17,3),(19,3),(20,3),(21,3),(22,3),(23,3),(24,3),(3,2),(8,2),(11,2),(16,2),(21,2),(24,2),(1,1),(3,1),(4,1),(6,1),(8,1),(9,1),(11,1),(12,1),(14,1),(16,1),(17,1),(19,1),(21,1),(22,1),(24,1)} {
    	\path [fill=darkgray] \pt rectangle +(1,1);
    }	
		
\end{tikzpicture}}

\def\tikzChristmasII
{\begin{tikzpicture}[xscale=0.24,yscale=0.24]

    \def\R{0.5774}

    \foreach \pt in {(0.5,16.454),(2.5,16.454),(3.5,16.454),(4.5,16.454),(5.5,16.454),(7.5,16.454),(8.5,16.454),(10.5,16.454),(12.5,16.454),(13.5,16.454),(15.5,16.454),(16.5,16.454),(17.5,16.454),(18.5,16.454),(20.5,16.454),(21.5,16.454),(23.5,16.454),(24.5,16.454),(4.0,15.588),(17.0,15.588),(1.5,14.722),(3.5,14.722),(4.5,14.722),(6.5,14.722),(9.5,14.722),(11.5,14.722),(14.5,14.722),(16.5,14.722),(17.5,14.722),(19.5,14.722),(22.5,14.722),(24.5,14.722),(1.0,13.856),(2.0,13.856),(3.0,13.856),(4.0,13.856),(5.0,13.856),(6.0,13.856),(7.0,13.856),(9.0,13.856),(10.0,13.856),(11.0,13.856),(12.0,13.856),(14.0,13.856),(15.0,13.856),(16.0,13.856),(17.0,13.856),(18.0,13.856),(19.0,13.856),(20.0,13.856),(22.0,13.856),(23.0,13.856),(24.0,13.856),(2.5,12.990),(5.5,12.990),(10.5,12.990),(15.5,12.990),(18.5,12.990),(23.5,12.990),(0.0,12.124),(2.0,12.124),(3.0,12.124),(5.0,12.124),(6.0,12.124),(8.0,12.124),(10.0,12.124),(11.0,12.124),(13.0,12.124),(15.0,12.124),(16.0,12.124),(18.0,12.124),(19.0,12.124),(21.0,12.124),(23.0,12.124),(24.0,12.124),(1.0,10.392),(4.0,10.392),(7.0,10.392),(9.0,10.392),(12.0,10.392),(14.0,10.392),(17.0,10.392),(20.0,10.392),(22.0,10.392),(0.5,9.526),(1.5,9.526),(3.5,9.526),(4.5,9.526),(6.5,9.526),(7.5,9.526),(8.5,9.526),(9.5,9.526),(11.5,9.526),(12.5,9.526),(13.5,9.526),(14.5,9.526),(16.5,9.526),(17.5,9.526),(19.5,9.526),(20.5,9.526),(21.5,9.526),(22.5,9.526),(24.5,9.526),(0.0,8.660),(8.0,8.660),(13.0,8.660),(21.0,8.660),(0.5,7.794),(2.5,7.794),(5.5,7.794),(7.5,7.794),(8.5,7.794),(10.5,7.794),(12.5,7.794),(13.5,7.794),(15.5,7.794),(18.5,7.794),(20.5,7.794),(21.5,7.794),(23.5,7.794),(0.0,6.928),(1.0,6.928),(2.0,6.928),(3.0,6.928),(5.0,6.928),(6.0,6.928),(7.0,6.928),(8.0,6.928),(9.0,6.928),(10.0,6.928),(11.0,6.928),(12.0,6.928),(13.0,6.928),(14.0,6.928),(15.0,6.928),(16.0,6.928),(18.0,6.928),(19.0,6.928),(20.0,6.928),(21.0,6.928),(22.0,6.928),(23.0,6.928),(24.0,6.928),(1.5,6.062),(6.5,6.062),(9.5,6.062),(11.5,6.062),(14.5,6.062),(19.5,6.062),(22.5,6.062),(24.5,6.062),(1.0,5.196),(2.0,5.196),(4.0,5.196),(6.0,5.196),(7.0,5.196),(9.0,5.196),(10.0,5.196),(11.0,5.196),(12.0,5.196),(14.0,5.196),(15.0,5.196),(17.0,5.196),(19.0,5.196),(20.0,5.196),(22.0,5.196),(23.0,5.196),(24.0,5.196),(10.5,4.330),(23.5,4.330),(0.0,3.464),(3.0,3.464),(5.0,3.464),(8.0,3.464),(10.0,3.464),(11.0,3.464),(13.0,3.464),(16.0,3.464),(18.0,3.464),(21.0,3.464),(23.0,3.464),(24.0,3.464),(0.5,2.598),(2.5,2.598),(3.5,2.598),(4.5,2.598),(5.5,2.598),(7.5,2.598),(8.5,2.598),(9.5,2.598),(10.5,2.598),(11.5,2.598),(12.5,2.598),(13.5,2.598),(15.5,2.598),(16.5,2.598),(17.5,2.598),(18.5,2.598),(20.5,2.598),(21.5,2.598),(22.5,2.598),(23.5,2.598),(24.5,2.598),(4.0,1.732),(9.0,1.732),(12.0,1.732),(17.0,1.732),(22.0,1.732),(1.5,0.866),(3.5,0.866),(4.5,0.866),(6.5,0.866),(8.5,0.866),(9.5,0.866),(11.5,0.866),(12.5,0.866),(14.5,0.866),(16.5,0.866),(17.5,0.866),(19.5,0.866),(21.5,0.866),(22.5,0.866),(24.5,0.866)} {
    	\draw [fill=darkgray,color=darkgray] \pt+(30:\R) -- +(90:\R) -- +(150:\R) -- +(210:\R) -- +(270:\R) -- +(330:\R) -- cycle;
    }
		
\end{tikzpicture}}

\usepackage[english]{babel}
\usepackage{csquotes}
\usepackage{amsmath,amssymb}
\usepackage{graphicx}
\usepackage{enumerate}
\usepackage{verbatim}

\usepackage[style=alphabetic,citestyle=alphabetic,sorting=nty,firstinits=true,backend=bibtex]{biblatex}
\bibliography{bibliography.bib}{}
\DeclareBibliographyCategory{excluded}

\usepackage[usenames,dvipsnames]{xcolor}
\usepackage{mathtools}
\usepackage{bbm} 
\usepackage{tikz}
\usetikzlibrary{calc}
\usepackage{subcaption}
\usepackage{hyperref}

\usepackage{xcolor}
\definecolor{dark-red}{rgb}{0.4,0.15,0.15}
\definecolor{dark-blue}{rgb}{0.15,0.15,0.4}
\definecolor{medium-blue}{rgb}{0,0,0.5}
\hypersetup{
  colorlinks,
  linkcolor={dark-red},
  citecolor={dark-blue},
  urlcolor={medium-blue}
}

\newcommand{\env}[2]{\begin{#1}#2\end{#1}}%
\newcommand{\envparam}[3][]{\begin{#2}[#1]#3\end{#2}}%

\spnewtheorem*{conjecture*}{Conjecture}{\itshape}{\rmfamily}
\spnewtheorem*{note*}{Note}{\itshape}{\rmfamily}

\def \N {\mathbb{N}}
\def \Z {\mathbb{Z}}
\def \Q {\mathbb{Q}}
\def \R {\mathbb{R}}
\def \C {\mathbb{C}}

\def \L {\mathcal{L}}
\def \u {{\vec{u}}}
\def \v {{\vec{v}}}
\def \w {{\vec{w}}}

\DeclarePairedDelimiter\abs{\lvert}{\rvert}
\DeclareMathOperator{\supp}{supp}
\DeclareMathOperator{\Ann}{Ann}
\DeclareMathOperator{\bbox}{box}
\DeclareMathOperator{\opc}{ord}%
\newcommand{\makeinner}[2]{ \, #1 \mid #2 \, }
\newcommand{\makeset}[2]{\{ \makeinner{#1}{#2} \}}
\newcommand{\makesetbig}[2]{\big\{ \, #1 \; \big| \; #2 \, \big\}}%
\newcommand{\gen}[1]{\langle #1 \rangle}

\title{An Algebraic Geometric Approach to Nivat's Conjecture}
\author{Jarkko Kari \and Michal Szabados}
\institute{
    Department of Mathematics and Statistics,\\
    University of Turku, 20014 Turku, Finland\\
    \href{mailto:jkari@utu.fi}{\tt jkari@utu.fi},
    \href{mailto:micsza@utu.fi}{\tt micsza@utu.fi} \\
    \ \\
    \small{\textit{Version: \today}}    
}

\begin{document}
\maketitle

\env{abstract}{We study  multidimensional configurations (infinite words) and subshifts of low pattern complexity using tools of algebraic geometry.
We express the configuration as a multivariate formal power series over integers and investigate the setup when there is a non-trivial annihilating polynomial: a non-zero polynomial whose formal product with the power series is zero. Such annihilator exists, for example, if the number of
distinct patterns of some finite shape $D$ in the configuration
is at most the size $|D|$ of the shape. This is our low pattern complexity assumption.
We prove that the configuration must be a sum of periodic configurations over integers, possibly with unbounded values. As a specific application of the method we obtain an asymptotic version of the well-known Nivat's conjecture: we prove that any two-dimensional, non-periodic configuration can satisfy the low pattern complexity assumption with respect to only finitely many distinct rectangular shapes $D$.
}
\keywords{
    Nivat's conjecture, symbolic dynamics, algebraic geometry, Laurent polynomials, pattern complexity, periodicity
}



\section{Introduction}

Consider configuration $c\in A^{\Z^d}$, a $d$-dimensional infinite array filled by symbols from finite alphabet $A$. Suppose that for some finite observation window $D\subseteq \Z^d$, the number of distinct patterns of shape $D$ that exist in $c$ is small, at most the cardinality $|D|$ of $D$. We investigate global regularities and structures in $c$ that are enforced by such local complexity assumption.

Let us be more precise on the involved concepts. As usual, we denote by $c_{\v} \in A$ the symbol in $c$ in position ${\v} \in\Z^d$. For $\u\in\Z^d$, $\u \ne 0$, we say that $c$ is {\em $\u$-periodic\/} if $c_\v=c_{\u+\v}$ holds for all $\v\in\Z^d$, and $c$ is {\em periodic\/} if it is $\u$-periodic for some $\u\neq 0$. For a finite domain $D\subseteq \Z^d$, the elements of $A^D$ are {\em $D$-patterns}. For a fixed $D$, we denote by $c_{\v+D}$ the $D$-pattern in $c$ in position $\v$, that is, the pattern $\u\mapsto c_{\v+\u}$ for all $\u\in D$. The number of distinct $D$-patterns in $c$ is the {\em $D$-pattern complexity\/} $P_c(D)$ of $c$. Our assumption of
low local complexity is
\begin{equation}
\label{eq:lowcomplexity}
P_c(D)\leq |D|,
\end{equation}
for some finite $D$.

\subsection*{Nivat's conjecture}

There are specific examples in the literature of open problems in this framework. {\em Nivat's conjecture\/} (proposed by M. Nivat in his keynote address in ICALP 1997~\cite{Nivat97}) claims that in the two-dimensional case $d=2$, the low complexity assumption (\ref{eq:lowcomplexity}) for a {\em rectangle} $D$ implies that $c$ is periodic.
The conjecture is a natural generalization of the one-dimensional Morse-Hedlund theorem that states that if a bi-infinite word contains at most $n$ distinct subwords of length $n$ then the word must be periodic~\cite{MorseHedlund38}.
In the two-dimensional setting and $m,n\in\N$ we denote by
$P_c(m,n)$ the complexity $P_c(D)$ for the $m\times n$ rectangle $D$.

\begin{conjecture}[Nivat's conjecture] \label{nivats-conjecture}
If for some $m,n$ we have $P_c(m,n)\leq mn$ then $c$ is periodic.
\end{conjecture}

The conjecture has recently raised wide interest, but it remains unsolved.
In~\cite{EpifanioKoskasMignosi03}  it was shown $P_c(m,n)\leq mn/144$ is enough to guarantee the periodicity of $c$. This bound was improved  to $P_c(m,n)\leq mn/16$ in~\cite{QuasZamboni04}, and
recently to $P_c(m,n)\leq mn/2$ in~\cite{CyrKra15}. Also the cases of narrow rectangles have been investigated: it was shown in~\cite{SanderTijdeman02} and recently in~\cite{CyrKra16} that $P_c(2,n)\leq 2n$ and $P_c(3,n)\leq 3n$, respectively, imply that $c$ is periodic. Note that it is enough to prove \autoref{nivats-conjecture} for two-letter alphabet (\autoref{lem:binary}).

The analogous conjecture in the higher dimensional setups $d\geq 3$ is false~\cite{SanderTijdeman00}. The following example recalls a simple counter example for $d=3$.

\begin{example}
\label{ex:3d}
Fix $n\geq 3$, and consider the following $c\in \{0,1\}^{\Z^3}$ consisting of two perpendicular lines of $1$'s on a $0$-background, at distance $n$ from each other: $c(i,0,0)=c(0,n,i)=1$ for all $i\in \Z$, and $c(i,j,k)=0$ otherwise (see \autoref{fig-ex-3d}). For $D$ equal to the $n\times n\times n$  cube we
have $P_c(D)=2n^2+1$ since the $D$-patterns in $c$ have at most a single $1$-line piercing a face of the cube. Clearly $c$ is not periodic although
$P_c(D)=2n^2+1<n^3=|D|$. Notice that $c$ is a ``sum'' of two periodic components (the lines of $1$'s). Our results imply that any counter example must decompose into a sum of periodic components.
\qed
\end{example}

\envparam{figure}{
    \centering
    \includegraphics[scale=0.7]{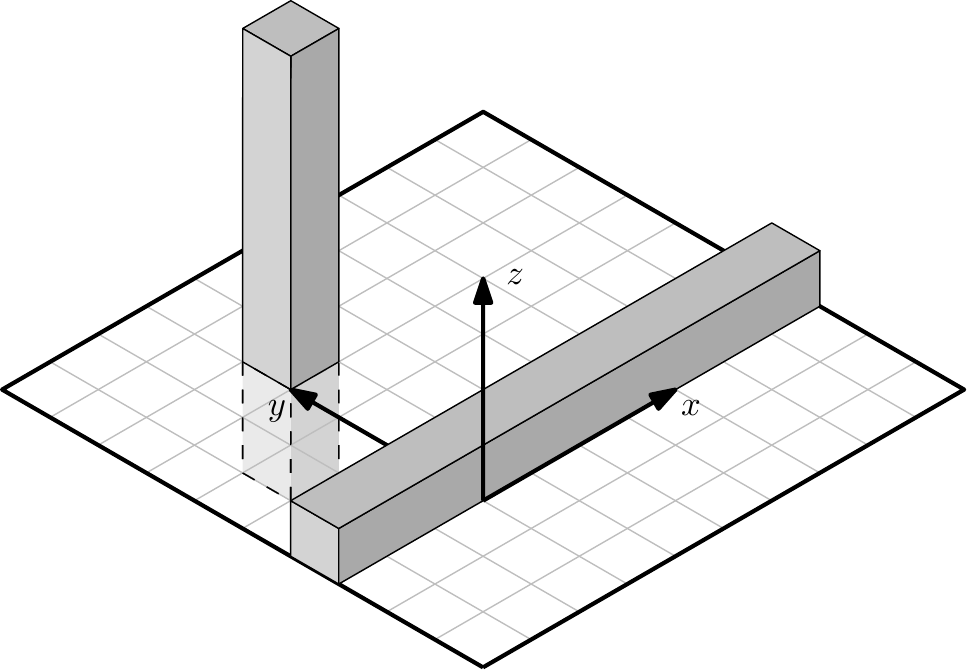}
    \caption{Non-periodic configuration of low complexity. If $D$ is the $4\times4\times4$ cube, then $ P(D) = 33$.}
    \label{fig-ex-3d}
}

\subsection*{Periodic tiling problem}

Another related open problem is the {\em periodic (cluster) tiling problem\/} by Lagarias and Wang~\cite{LagariasWang}.
A (cluster) tile is a finite  $D\subset \Z^d$. Its co-tiler is any subset $C\subseteq \Z^d$ such that
\begin{equation}
\label{eq:tiling}
D\oplus C = \Z^d.
\end{equation}
The co-tiler can be interpreted as the set of positions where copies of $D$ are placed so that they together cover the entire $\Z^d$ without overlaps. Note that the tile $D$ does not need to be connected -- hence the term
``cluster tile'' is sometimes used. The tiling is by translations of $D$ only: the tiles may not be rotated.

It is natural to interpret any $C\subseteq \Z^d$ as the binary
configuration $c\in \{0,1\}^{\Z^d}$ with $c_\v=1$ if and only if $\v\in C$.
Then the tiling condition (\ref{eq:tiling}) states that $C$ is a co-tiler for $D$ if and only if
the ($-D$)-patterns in the corresponding configuration $c$ contain exactly a single $1$ in the background of $0$'s. In fact, as co-tilers of $D$ and $-D$ coincide~\cite{Szegedy}, this is equivalent to all $D$-patterns having a single $1$.

We see that the set ${\cal C}$ of all co-tiler configurations
for $D$ is a {\em subshift of finite type\/}~\cite{LindMarcus}.
We also see that the low local complexity assumption (\ref{eq:lowcomplexity}) is satisfied. We even have $P_{\cal C}(D)\leq |D|$ where we denote by $P_{\cal C}(D)$ the number of distinct
$D$-patterns found in the elements of the subshift ${\cal C}$.

\begin{conjecture}[Periodic Tiling Problem]
If tile $D$ has a co-tiler then it has a periodic co-tiler.
\end{conjecture}

This conjecture was first formulated in~\cite{LagariasWang}. In the one-dimensional case it is easily seen true. The two-dimensional case was established only recently \cite{Bhattacharya}, the higher dimensional cases with $d>2$ are open. Interestingly, it is known that if $|D|$ is a prime number then {\it every\/} co-tiler of $D$ is periodic~\cite{Szegedy} (see also our Example~\ref{ex:prime}).

\subsection*{Our contributions}

We approach these problems using tools of algebraic geometry. Assuming alphabet
$A\subseteq\Z$, we express configuration $c$ as a formal power series over $d$ variables and with coefficients in $A$. The complexity assumption (\ref{eq:lowcomplexity}) implies that there is a non-trivial polynomial that annihilates the power series under formal multiplication (\autoref{lem-low-cplx-has-annihilator}). This naturally leads to the study of the annihilator ideal of the power series, containing all the polynomials that annihilate it. Using Hilbert's Nullstellensatz we prove that the ideal contains polynomials of particularly simple form (\autoref{thm-product-of-differences}). In particular, this implies that $c=c_1+\dots +c_m$ for some periodic $c_1,\dots ,c_m$ (\autoref{thm-decomposition-theorem}). This decomposition result is already an interesting global structure on $c$, but to prove periodicity we would need $m=1$.

We study the structure of the annihilator ideal in the two-dimensional setup, and prove that it is always a radical (\autoref{ann-is-radical}). This leads to a stronger decomposition theorem (\autoref{thm-2d-decomposition}).

To approach Nivat's conjecture we study a hypothetical non-periodic configuration that would be a counterexample to it. Our main result is an asymptotic version of the conjecture (\autoref{thm-main-result}): for any non-periodic configuration $c$ there are only finitely many pairs $m,n\in\N$ such that $P_c(m,n)\leq mn$. 

These results were reported without detailed proofs at ICALP 2015 conference \cite{KariSzabados2015ICALP}.

\section{Basic Concepts and Notation}
\label{sec:basics}

For a domain $R$ -- which will usually be the whole numbers $\Z$ or complex numbers $\C$ -- denote by $R[x_1,\dots,x_d]$ the set of polynomials over $R$ in $d$ variables. We adopt the usual simplified notation: for a $d$-tuple of non-negative integers $\v = (v_1,\dots,v_d)$ set $X^\v = x_1^{v_1}\dots x_d^{v_d}$, then we write
\env{align*}{
    R[X] = R[x_1,\dots,x_d]
}
and a general polynomial $f \in R[X]$ can be expressed as $f = \sum a_\v X^\v$, where $a_\v \in R$ and the sum goes over finitely many $d$-tuples of non-negative integers $\v$. If we allow $\v$ to contain also negative integers we obtain \emph{Laurent polynomials}, which are denoted by $R[X^{\pm1}]$. Finally, by relaxing the requirement to have only finitely many $a_\v \ne 0$ we get \emph{formal power series}:
\env{align*}{
    R[[X^{\pm1}]] = \makesetbig{\!\sum a_\v X^\v}{\v \in \Z^d,\ a_\v \in R}.
}
Note that we allow infinitely many negative exponents in formal power series.

\bigskip

Let $d$ be a positive integer. Let us define a $d$-dimensional \emph{configuration} to be any formal power series $c \in \C[[X^{\pm1}]]$ and denote by $c_\v$ the coefficient of $X^\v$:
\env{align*}{
    c = \sum_{\v \in \Z^d} c_\v X^\v
}
A configuration is \emph{integral} if all coefficients $c_{\vec v}$ are integers, and it is \emph{finitary} if there are only finitely many distinct coefficients $c_{\vec v}$.

Classically in symbolic dynamics configurations are understood as elements of $A^{\Z^d}$. Because the actual names of the symbols in the alphabet $A$ do not matter, they can be chosen to be integers. Then such a ``classical'' configuration can be identified with a finitary integral configuration by simply setting the coefficient $c_\v$ to be the integer at position $\v$.

Multiplication of a formal power series by a Laurent polynomial is well defined and results again in formal power series. For example, $X^\v c$ is a translation of $c$ by the vector $\v$. Another important example is that $c$ is periodic if and only if there is a non-zero $\v \in \Z^d$ such that $(X^\v-1)c = 0$. Here the right side is understood as the constant zero configuration.

For a polynomial $f(X) = \sum a_\v X^\v$ and a positive integer $n$ define $f(X^n) = \sum a_\v X^{n\v}$. (See \autoref{fig:scaled_polynomial}.) The following example, and the proof of Lemma~\ref{lem-expanding-annihilator}, use the well known fact that
for any integral polynomial $f$ and prime number $p$, we have $f^p(X) \equiv f(X^p) \pmod{p}$.

\env{figure}{
    \centering
    \includegraphics[scale=0.9]{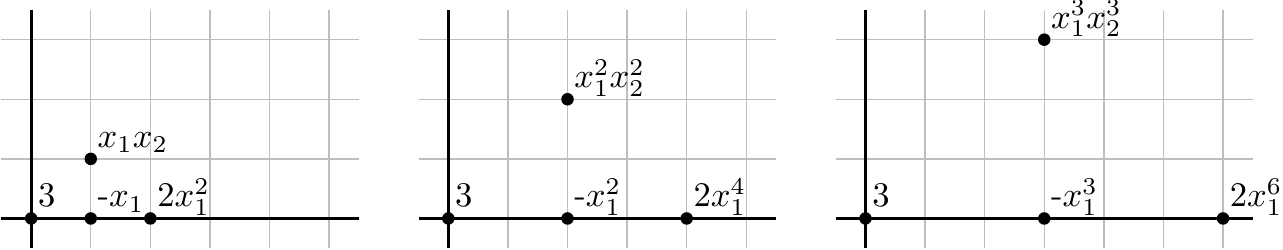}
    \caption{Plot of $f(X)$, $f(X^2)$ and $f(X^3)$ for the polynomial $f(X) = f(x_1,x_2) = 3 - x_1 + 2 x_1^2 + x_1x_2$.}
    \label{fig:scaled_polynomial}
}

\begin{example}
\label{ex:prime}
The example concerns the periodic tiling problem. We
provide a short proof of the fact -- originally proved in~\cite{Szegedy} -- that if the size $p=|D|$ of tile $D$ is a prime number then all co-tilers $C$ are periodic. When the tile $D$
is represented as the Laurent polynomial $f(X)=\sum_{\v \in D} X^\v$ and the co-tiler $C$
as the power series $c(X)=\sum_{\v \in C} X^\v$, the tiling condition (\ref{eq:tiling})
states that $f(X)c(X)=\sum_{\v \in \Z^d} X^\v$. Multiplying both sides by $f^{p-1}(X)$, we get
$$f^p(X)c(X)=\sum_{\v \in \Z^d} p^{p-1}X^\v\equiv 0\pmod{p}.$$
On the other hand, since $p$ is a prime, $f^p(X) \equiv f(X^p) \pmod{p}$ so that 
$$f(X^p)c(X)\equiv 0\pmod{p}.$$
Let $\v\in D$ and $\vec{w}\in C$ be arbitrary. We have
$$0\equiv [f(X^p)c(X)]_{\vec{w}+p\v} = \sum_{\u\in D} c(X)_{\vec{w}+p\v-p\u}\pmod{p}.$$
The last sum is a sum of $p$ numbers, each $0$ or $1$, among which there is at least one $1$ (corresponding to $\u=\v$). The only way for the sum to be divisible by $p$  is by having each summand equal to $1$. We have that $\vec{w}+p(\v-\u)$ is in $C$ for all $\u,\v\in D$ and $\vec{w}\in C$, which means
that $C$ is $p(\v-\u)$-periodic for all $\u,\v\in D$.\qed

\end{example}

The next lemma grants us that for low complexity configurations there exists at least one
Laurent polynomial that annihilates the configuration by formal multiplication.

\env{lemma}{
    \label{lem-low-cplx-has-annihilator}
    Let $c$ be a configuration and $D \subset \Z^d$ a finite domain such that $P_c(D) \leq |D|$. Then there exists a non-zero Laurent polynomial $f \in \C[X^{\pm1}]$ such that $fc=0$.
}
\env{proof}{
    Denote $D = \{\vec{u_1},\dots,\vec{u_n}\}$ and consider the set
    \env{align*}{
        \makeset{(1,c_{\vec{u_1}+\v},\dots,c_{\vec{u_n}+\v})}{\v \in \Z^d}.
    }
    It is a set of complex vectors of dimension $n + 1$, and because $c$ has low complexity there is at most $n$ of them. Therefore there exists a common non-zero orthogonal vector $(\overline{a_0},\dots,\overline{a_n})$. Let $g(X) = a_1X^{-\vec{u_1}} + \dots + a_nX^{-\vec{u_n}} \ne 0$, then the coefficient of $gc$ at position $\v$ is
    \env{align*}{
        (gc)_\v = a_1 c_{\vec{u_1}+\v} + \dots + a_n c_{\vec{u_n}+\v} = -a_0,
    }
    that is, $gc$ is a constant configuration. Now it suffices to set $f = (X^\v-1)g$ for arbitrary non-zero vector $\v \in \Z^d$.
\qed}

\section{Annihilating Polynomials}
\label{sec-annihilators}

Let $c$ be a configuration. We say that a Laurent polynomial $f$ \emph{annihilates} (or is an \emph{annihilator} of) the configuration if $fc=0$. Define
\env{align*}{
    \Ann(c) = \makesetbig{f \in \C[X]}{fc = 0}.
}
It is the set of all polynomial annihilators of $c$. Clearly it is an ideal of $\C[X]$. The zero polynomial annihilates every configuration; let us call the annihilator \emph{non-trivial} if it is non-zero.

An easy, but useful observation is that if $f$ is an annihilator, then any monomial multiple $X^\v f$ is also an annihilator. We shall use this fact without further reference.

There is a good reason why to study this ideal. Firstly, by \autoref{lem-low-cplx-has-annihilator}, for low complexity configurations $\Ann(c)$ is non-trivial, which is the case of Nivat's conjecture and periodic tiling problem. Secondly, to prove that a configuration is periodic is equivalent to showing that $X^\v - 1$ annihilates $c$ for some non-zero $\v \in \Z^d$.

We defined $\Ann(c)$ to consist of complex polynomials, so that we can later use Hilbert's Nullstellensatz directly, as it requires polynomial ideals over algebraically closed field. We shall however occasionally work with integer coefficients and Laurent polynomials when it is more convenient.

In what follows we consider configurations which have an integral annihilator. Although it follows by a small modification of \autoref{lem-low-cplx-has-annihilator} that such an annihilator for integral configurations exists, a stronger statement holds:

\env{lemma}{
    \label{lem-integral-generators}
	Let $c$ be an integral configuration. Then $\Ann(c)$ is generated by finitely many integral polynomials.
}
\env{proof}{
    We will show that $\Ann(c)$ is generated by integral polynomials, the claim then follows from Hilbert's Basis Theorem. Let $f \in \Ann(c)$ be arbitrary and denote
    \[
        f(X) = \sum_{i=1}^n a_i X^{\vec{u_i}}.
    \]	
	Let $V$ be a vector subspace of $\C^n$ defined by
	\[
	    V := \left\langle \makeinner{(c_{\v-\vec{u_1}},\dots,c_{\v-\vec{u_n}})}{\v \in \Z^d} \right\rangle.
	\]
	Then $fc=0$ if and only if $(\overline{a_1},\dots,\overline{a_n}) \perp V$. All the vectors in $V$ have integers coordinates, therefore the space $V^\perp$ has a basis consisting of rational, and therefore also integer vectors $\vec{b}^{(1)}, \dots, \vec{b}^{(m)}$. Denote $\vec{b}^{(j)}=(b_1^{(j)},\dots,b_n^{(j)})$.

	Consider integral polynomials $g^{(j)}(X) = \sum_{i=1}^n b_i^{(j)}X^{\vec{u_i}}$. Because $\overline{\vec{b}^{(j)}} = \vec{b}^{(j)} \perp V$ we have that $g^{(j)}$ is an integral annihilator of $c$. From construction the polynomial $f$ is a linear combination of $g^{(1)},\dots,g^{(m)}$, which concludes the proof.
\qed
}

In this section we aim to prove a decomposition theorem -- the fact that every finitary integral configuration with an annihilator can be written as a sum of periodic configurations. Let us introduce additional notation: if $Z = (z_1, \dots, z_d) \in \C^d$ is a complex vector, then it can be plugged into a polynomial. In particular, plugging into a monomial $X^\v$ results in $Z^\v = z_1^{v_1} \cdots z_d^{v_d}$. Recall that the notation $f(X^n)$ for positive integers $n$ was defined in \autoref{sec:basics}.

\env{lemma}{
    \label{lem-expanding-annihilator}
    Let $c(X)$ be a finitary integral configuration and $f(X) \in \Ann(c)$ a non-zero integer polynomial. Then there exists an integer $r$ such that for every positive integer $n$ relatively prime to $r$ we have $f(X^n) \in \Ann(c)$.
}
\env{proof}{
    Denote $f(X) = \sum a_\v X^\v$ and let $m \in \N$ be arbitrary. We prove that if $f(X^m)$ is an annihilator, then also $f(X^{pm})$ is an annihilator for a large enough prime $p$.

    Let $p$ be a prime. Since $f^p(X) \equiv f(X^p) \pmod{p}$ we especially have $f^p(X^m) \equiv f(X^{pm}) \pmod{p}$. We assume that $f(X^m)$ annihilates $c(X)$, therefore multiplying both sides by $c(X)$ results in
    \env{align*}{
        0 \equiv f(X^{pm})c(X) \pmod{p}.
    }
    The coefficients in $f(X^{pm})c(X)$ are bounded in absolute value by
    \env{align*}{
        s = c_{max}\sum \abs {a_\v},
    }
    where $c_{max}$ is the maximum absolute value of coefficients in $c$. Note that the bound is independent of $m$. Therefore for any $m$, if $p > s$ we have $f(X^{pm})c(X) = 0$, which means $f(X^{pm}) \in \Ann(c)$.

    To finish the proof, set $r = s!$. Now every $n$ relatively prime to $r$ is of the form $p_1\cdots p_k$ where each $p_i$ is a prime greater than $s$. Because $f(X)$ is an annihilator now it follows easily by induction that also $f(X^{p_1\cdots p_k})$ is an annihilator.
\qed}

 Let us define the \emph{support} of a Laurent polynomial $f = \sum a_\v X^\v$ as
 \env{align*}{
 	\supp(f) = \makeset{\v \in \Z^d}{a_\v \ne 0}.
 }
Recall that $x_1,\dots,x_d$ denote the variables of polynomials.

\env{lemma}{
    \label{lem-polynomial-from-radical}
    Let $c$ be a finitary integral configuration and $f = \sum a_\v X^\v$ a non-trivial integer polynomial annihilator. Define
    \env{align*}{
        g(X) = x_1\cdots x_d\prod_{\substack{\v\in \supp(f)\\ \v \ne \v_0}}\left(X^{r\v}-X^{r\vec{v_0}}\right)
    }
    where $r$ is the integer from \autoref{lem-expanding-annihilator} and $\vec{v_0} \in \supp(f)$ arbitrary. Then $g(Z) = 0$ for any common root $Z \in \C^d$ of $\Ann(c)$.
}
\env{proof}{
    Fix $Z$. If any of its complex coordinates is zero then clearly $g(Z) = 0$. Assume therefore that all coordinates of $Z$ are non-zero.

    Let us define for $\alpha \in \C$
    \env{align*}{
        S_\alpha &= \makesetbig{\v \in \supp(f)}{Z^{r\v} = \alpha}, \\
        f_\alpha(X) &= \sum_{\v \in S_\alpha} a_\v X^\v.
    }
    Because $\supp(f)$ is finite, there are only finitely many non-empty sets $S_{\alpha_1}, \dots, S_{\alpha_m}$ and they form a partitioning of $\supp(f)$. In particular we have $f = f_{\alpha_1} + \dots + f_{\alpha_m}$.

    Numbers of the form $1+ir$ are relatively prime to $r$ for all non-negative integers $i$, therefore by \autoref{lem-expanding-annihilator}, $f(X^{1+ir}) \in \Ann(c)$. Plugging in $Z$ we obtain $f(Z^{1+ir}) = 0$. Now compute:
    \env{align*}{
        f_\alpha(Z^{1+ir}) &= \sum_{\v \in S_\alpha} a_\v Z^{(1+ir)\v} = \sum_{\v \in S_\alpha} a_\v Z^\v \alpha^i = f_\alpha(Z)\alpha^i}
    Summing over $\alpha=\alpha_1,\dots, \alpha_m$ gives
     \env{align*}{
        0 = f(Z^{1+ir}) &= f_{\alpha_1}(Z)\alpha_1^i + \dots + f_{\alpha_m}(Z)\alpha_m^i
    }
    Let us rewrite the last equation as a statement about orthogonality of two vectors in $\C^m$:
    \env{align*}{
        \left( \overline{f_{\alpha_1}(Z)}, \dots, \overline{f_{\alpha_m}(Z)} \right) \perp (\alpha_1^i, \dots, \alpha_m^i)
    }
    By Vandermode determinant, for $i \in \{0,\dots,m-1\}$ the vectors on the right side span the whole $\C^m$. Therefore the left side must be the zero vector, and especially for $\alpha$ such that $\vec{v_0} \in S_\alpha$ we have
    \env{align*}{
        0 = f_{\alpha}(Z) = \sum_{\v \in S_{\alpha}} a_\v Z^\v.
    }
    Because $Z$ does not have zero coordinates, each term on the right hand side is non-zero. But the sum is zero, therefore there are at least two vectors $\vec{v_0},\v \in S_\alpha$. From the definition of $S_\alpha$ we have $Z^{r\v} = Z^{r\vec{v_0}} = \alpha$, so $Z$ is a root of $X^{r\v}-X^{r\vec{v_0}}$.
\qed}

\subsection*{Line polynomials}

We say that a Laurent polynomial $f$ is a \emph{line Laurent polynomial} if its support contains at least two points and all the points lie on a single line. Let us call a vector $\v \in \Z^d$ \emph{primitive} if its coordinates don't have a common non-trivial integer factor. Then every line Laurent polynomial can be expressed as
\env{align*}{
    f(X) = X^{\v'}(a_n X^{n\v} + \dots + a_1 X^\v + a_0)
}
for some $a_i \in \C$, $n\geq 1$, $a_n \ne 0 \ne a_0$, $\v',\v \in \Z^d$, where $\v$ is primitive. Moreover, the vector $\v$ is determined uniquely up to the sign. We define the \emph{direction} of a line Laurent polynomial to be the vector space $\gen \v \subset \Q^d$.

Recall that an ideal $A \leq \C[X]$ is \emph{radical} if $a^n \in A$ implies $a \in A$. Clearly, that happens if and only if $A = \sqrt A$ where
\env{align*}{
    \sqrt A = \makesetbig{a \in \C[X]}{\exists n: a^n \in A}.
}
The next lemma states that for one-dimensional configurations $\Ann(c)$ is radical.

\env{lemma}{
    \label{lem-ann-1d-radical}
	Let $c \in \C[[x^{\pm1}]]$ be a finitary one-dimensional configuration annihilated by $f^m$ for a non-trivial polynomial $f$ and $m \in \N$. Then it is also annihilated by $f$.
}
\env{proof}{
    The configuration $c$ can be viewed as a sequence attaining only finitely many values, and $f^m$ as a recurrence relation on it. Therefore $c$ must be periodic, which means there is $n \in \N$ such that $x^n-1 \in \Ann(c)$.

		Then also $g = \gcd(x^n-1,f^m) \in \Ann(c)$. Because $g$ divides $x^n-1$, it has only simple roots, and from $g \mid f^m$ we conclude $g \mid f$. Any multiple of $g$ annihilates the sequence, hence also $f$ does.
\qed}

\env{lemma}{
    \label{lem-prod-of-1d-radicality}
    Let $c$ be a finitary configuration and $f_1, \dots, f_k$ line Laurent polynomials such that $f_1^{m_1}\cdots f_k^{m_k}$ annihilates $c$. Then also $f_1\cdots f_k$ annihilates it.
}
\env{proof}{
    We will show that if $f$ is a line Laurent polynomial and $f^m$ annihilates $c$, then also $f$ annihilates $c$. Without loss of generality assume
    \env{align*}{
	    f(X) = a_nX^{n\v} + \dots + a_1X^\v + a_0
    }
    for some $a_i \in \C$ and $\v \in \Z^d$. Define $g(t) = a_nt^n + \dots + a_1t + a_0 \in \C[t]$ so that $f^m(X) = g^m(X^\v)$.
    
    For any $\u \in \Z^d$ the sequence of coefficients $(c_{\u+i\v})_{i \in \Z}$ can be viewed as a one-dimensional configuration annihilated by $g^m$. By \autoref{lem-ann-1d-radical} it is also annihilated by $g$, therefore $g(X^\v) = f(X)$ annihilates $c$.

    To finish the proof observe that $f_2^{m_2} \dots f_k^{m_k}c$ is a finitary configuration annihilated by $f_1^{m_1}$. Thus it is also annihilated by $f_1$ and $f_1f_2^{m_2} \dots f_k^{m_k}c = 0$. The argument can be repeated for all $f_i$.
\qed}

\env{theorem}{
    \label{thm-annihilator-from-support}
    Let $c$ be a finitary integral configuration and $f = \sum a_\v X^\v$ a non-trivial integral polynomial annihilator. Let $r$ be the integer from \autoref{lem-expanding-annihilator} and $\vec{v_0} \in \supp(f)$ arbitrary. Then the Laurent polynomial
    \env{align*}{
        \prod_{\substack{\v\in \supp(f)\\ \v \ne \v_0}}\left(X^{r\v}-X^{r\vec{v_0}}\right)
    }
    annihilates the configuration.
}
\env{proof}{
    Denote $g(X)$ the polynomial in the statement. By \autoref{lem-polynomial-from-radical}, $x_1 \cdots x_d \cdot g(X)$ vanishes on all common roots of $\Ann(c)$, therefore by Hilbert's nullstellensatz $x_1 \cdots x_d \cdot g(X) \in \sqrt{\Ann(c)}$. There exists an integer $m$ such that $x_1^m \cdots x_d^m \cdot g^m(X) \in \Ann(c)$. Then also $g^m(X)$ is an annihilator and the proof is finished by \autoref{lem-prod-of-1d-radicality}.
\qed}

\env{corollary}{
    \label{thm-product-of-differences}
    Let $c$ be a finitary integral configuration with a non-trivial annihilator. Then there exist vectors $\vec{v_1}, \dots, \vec{v_m} \in \Z^d$ in pairwise distinct directions such that the Laurent polynomial
    \env{align*}{
        (X^{\vec{v_1}}-1)\cdots (X^{\vec{v_m}}-1)
    }
    annihilates $c$.
}
\env{proof}{
    By \autoref{lem-integral-generators}, $c$ has an integral annihilating polynomial, and therefore also an annihilating polynomial as in \autoref{thm-annihilator-from-support}. Divide it by $X^{(\abs{\supp(f)}-1)r\vec{v_0}}$ to obtain an annihilator of the form $\prod (X^{\vec{u_i}}-1)$. To finish the proof observe that   $(X^{a\u}-1)(X^{b\u}-1)$ divides $(X^{ab\u}-1)^2$, and therefore any two factors $(X^{a\u}-1)(X^{b\u}-1)$ can be by \autoref{lem-prod-of-1d-radicality} replaced by a single factor $(X^{ab\u}-1)$.
\qed}

\subsection*{Decomposition theorem}

Multiplying a configuration by $(X^\v-1)$ can be seen as a \emph{"difference operator"} on the configuration. \autoref{thm-product-of-differences} then says, that there is a sequence of difference operators which annihilates the configuration. We can reverse the process: let us start by a zero configuration and step by step \emph{"integrate"} until we obtain the original configuration. This idea gives the Decomposition theorem:

\envparam[Decomposition theorem]{theorem}{
    \label{thm-decomposition-theorem}
    Let $c$ be a finitary integral configuration with a non-trivial annihilator. Then there exist periodic integral configurations $c_1, \dots, c_m$ such that $c = c_1 + \dots + c_m$.
}

The proof goes by a series of lemmas.

\env{lemma}{
    \label{lem-discrete-integration}
    Let $f,g$ be line Laurent polynomials in distinct directions and $c$ a configuration annihilated by $g$. Then there exists a configuration $c'$ such that $fc' = c$ and $c'$ is also annihilated by $g$.
}

\env{proof}{
    Without loss of generality assume $f,g$ are of the form
    \env{align*}{
        f(X) &= a_nX^{n\u} + \dots + a_1X^\u + a_0 \\
        g(X) &= b_mX^{m\v} + \dots + b_1X^\v + b_0
    }
    for some vectors $\u,\v \in \Z^d$, $n, m \in \N$ and $a_i, b_i \in \C$ such that $a_n, b_m, a_0, b_0$ are all non-zero.

    The vectors $\u$ and $\v$ are linearly independent and the whole space $\Z^d$ is partitioned into two-dimensional sublattices (cosets) modulo $\gen{\u,\v}$. Fix one such a sublattice $\Lambda$ and a point $\vec{z} \in \Lambda$, then every point in the sublattice can be uniquely expressed as $\vec{z} + a\u + b\v$ for some $a, b \in \Z$. Denote $[a,b] = \vec{z} + a\u + b\v$.

    The equation $fc' = c$ is satisfied if and only if
    \env{align}{
        a_n c'_{[a-n,b]} + \dots + a_1 c'_{[a-1,b]} + a_0 c'_{[a,b]} = c_{[a,b]} \label{eq-a}
    }
    holds for every $a,b \in\Z$ (on every sublattice $\Lambda$). This is a linear recurrence relation on the sequences $(c'_{[a,b]})_{a \in \Z}$. Let us define $c'_{[a,b]} = 0$ if $0 \leq a < n$, the rest of $c'$ is then uniquely determined by the recurrence relation so that $fc' = c$ holds.

    It remains to show that $c'$ defined this way is annihilated by $g$. A simple computation shows that
    \env{align*}{
    	f(gc') = g(fc') = gc = 0.
    }
    Therefore the configuration $gc'$ satisfies a linear recurring relation defined by $f$ on the sequences $\left((gc')_{[a,b]}\right)_{a \in \Z}$. Moreover we have $(gc')_{[a,b]} = 0$ for $0 \leq a < n$, from which it follows that $gc'$ is zero everywhere.
\qed}

\env{lemma}{
    \label{lem-decomposition-with-polynomials}
    Let $f_1,\dots,f_m$ be line Laurent polynomials in pairwise distinct directions and $c$ a configuration annihilated by their product $f_1 \cdots f_m$. Then there exist configurations $c_1, \dots, c_m$ such that $f_i$ annihilates $c_i$ and
    \env{align*}{
        c = c_1 + \dots + c_m.
    }
}

\env{proof}{
    The proof goes by induction on $m$. For $m=1$ there is nothing to prove, assume $m \geq 2$.

    Since the configuration $f_m c$ is annihilated by $f_1\cdots f_{m-1}$, by induction hypothesis we have
    \env{align*}{
        f_m c = b_1 + \dots + b_{m-1}
    }
    where each $b_i$ is annihilated by $f_i$ for $1 \leq i < m$. Let $c_i$ be such that $f_m c_i = b_i$ and $c_i$ is annihilated by $f_i$, this is possible by \autoref{lem-discrete-integration}. Then it suffices to set $c_m = c - c_1 - \dots - c_{m-1}$; clearly $c = c_1 + \dots + c_m$ and
    \env{align*}{
        f_m c_m = f_m (c - c_1 - \dots - c_{m-1}) = 0.
    }
\qed}

\envparam[of \autoref{thm-decomposition-theorem}]{proof}{
    By \autoref{thm-product-of-differences} there is an annihilator of the form $(X^{\vec{v_1}}-1)\cdots(X^{\vec{v_m}}-1)$ where $(X^{\vec{v_i}}-1)$ have distinct directions. Therefore by \autoref{lem-decomposition-with-polynomials} there are $c_1, \dots, c_m$ such that $c$ is their sum and each $c_i$ is periodic with the vector $\vec{v_i}$.

     It remains to show that $c_i$ can be integral. This follows from the fact that configurations in the proof of \autoref{lem-discrete-integration} are constructed by satisfying a recurrence relation (\ref{eq-a}), which for polynomials of the form $(X^{\vec{v_i}}-1)$ has always integral solution.
\qed}

\begin{example}
\label{ex:3d_again}
Recall the 3D counter example in \autoref{ex:3d}. It is the sum
$c_1+c_2$
where $c_1(i,0,0)=1$ and $c_2(0,n,i)=1$ for all $i\in \Z$, and all other entries
are $0$. Configurations $c_1$ and $c_2$ are $(1,0,0)$- and $(0,0,1)$-periodic, respectively, so that
$(X^{(1,0,0)}-1)(X^{(0,0,1)}-1)$ annihilates $c=c_1+c_2$.\qed
\end{example}

\begin{figure}
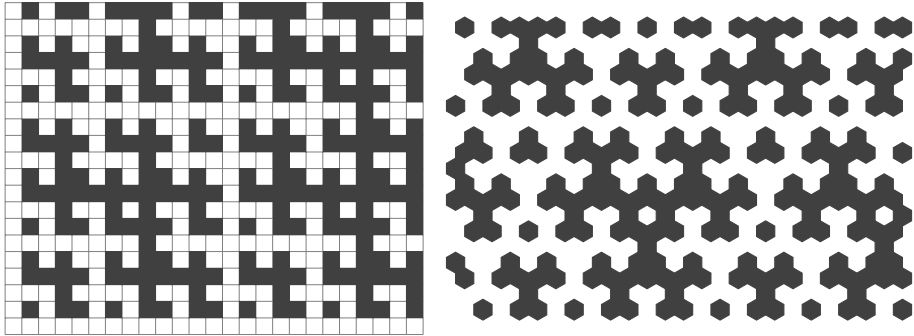

\centering
\begin{subfigure}{.5\textwidth}
  \centering
  \tikzChristmasI
\end{subfigure}%
\begin{subfigure}{.5\textwidth}
  \centering
  \tikzChristmasII
\end{subfigure}
\caption{The configuration $c$ from \autoref{ex:infinitary} when $\alpha$ is the golden ratio is shown on the left. On the right the configuration is skewed such that the three directions $\gen{(1,0)}$, $\gen{(0,1)}$ and $\gen{(1,-1)}$ became symmetrical, the bottom left corner is preserved.}
\label{fig:snowflakes}
\end{figure}

\begin{example}
\label{ex:infinitary}
The periodic configurations $c_1,\dots ,c_m$ in \autoref{thm-decomposition-theorem} may, for some configurations $c$, be necessarily non-finitary. Let $\alpha\in \R$ be irrational, and define three periodic two-dimensional configurations $c_1, c_2$ and $c_3$ by
$$
c_1(i,j)=\lfloor i\alpha\rfloor, \hspace*{1cm}
c_2(i,j)=\lfloor j\alpha\rfloor, \hspace*{1cm}
c_3(i,j)=\lfloor (i+j)\alpha\rfloor.
$$
Then $c=c_3-c_1-c_2$ is a finitary integral configuration (over alphabet $\{0,1\}$),
annihilated by the polynomial $(X^{(1,0)}-1)(X^{(0,1)}-1)(X^{(1,-1)}-1)$, but it cannot be expressed as a sum of finitary periodic configurations as proved in \cite{KariSzabados2015CAI}. \autoref{fig:snowflakes} illustrates the setup for $\alpha$ being the golden ratio.
\qed
\end{example}

\section{Two-dimensional Configurations}
\label{sec-ann-structure}

In the rest of the paper we focus on two-dimensional configurations. We analyze $\Ann(c)$ using tools of algebraic geometry and provide a description of a polynomial $\phi$ which divides every annihilator. Moreover we show a theoretical result that $\Ann(c)$ is a radical ideal, which allows us to formulate a more explicit version of the decomposition theorem for two-dimensional configurations.

To simplify the notation, we prefer to write $\C[x,y]$ in the place of $\C[x_1,x_2]$.
Let us recall some algebraic notions about polynomial ideals, for a reference see \cite{AtiyahMacDonald} and \cite{Cox}. By \emph{roots} or \emph{zeros} of an ideal $A \leq \C[X]$ we understand the set
\env{align*}{
    \makeset{Z \in \C^d}{\forall f \in A \colon f(Z) = 0}.
}
Two ideals $A, B \leq \C[X]$ are said to be \emph{comaximal} if $A + B = \C[X]$, or equivalently if $1 \in A + B$. It is a fact that two polynomial ideals in $\C[X]$ are comaximal if and only if they do not have common zeros. It is also a well-known fact that if $A_1, \dots, A_n$ are pairwise comaximal ideals then $\bigcap A_i = \prod A_i$.

Recall that an ideal $A$ is \emph{prime} if $ab \in A$ implies $a \in A$ or $b \in A$. We make use of the well-known minimal decomposition theorem for radical ideals and adapt it to the ring $\C[x,y]$.

\envparam[Minimal decomposition]{theorem}{
    \label{minimal-decomposition}
    Every radical ideal $A \leq \C[X]$ can be uniquely written as a finite intersection of prime ideals $A = P_1 \cap \dots \cap P_k$ where $P_i \not \subset P_j$ for $i \ne j$.
}
\env{proof}{
	 See e.g. \cite{Cox} Chapter 4, \S 6, Theorem 5.
\qed}

\env{lemma}{
    \label{primes-of-cxy}
    For a non-trivial prime ideal $P \leq \C[x,y]$ one of the following holds:
    \env{itemize}{
        \item $P$ is a principal ideal generated by an irreducible polynomial, i.e. $P = \gen \varphi$ for some irreducible $\varphi$,
        \item or $P$ is maximal ideal, in which case $P = \gen{ x-\alpha, y-\beta }$ for some $\alpha,\beta \in \C$.
    }
}
\env{proof}{
	Follows by Proposition 1 in section 1.5 and Corollary 2 in section 1.6 of Fulton's book \cite{Fulton}.
\qed}

Let us define the empty intersection and empty product of ideals to be the whole ring $\C[x,y]$.

\env{corollary}{
    \label{radical-decomposition-for-cxy}
    Let $A \leq \C[x,y]$ be a non-trivial radical ideal. Then there are distinct principal ideals $R_1, \dots, R_s$ generated by irreducible polynomials and distinct maximal ideals $M_1, \dots, M_t$ such that $R_i \not \subset M_j$ and
        $$A = R_1 \cdots R_s \, M_1 \cdots M_t.$$
        Moreover the ideals are determined uniquely and the ideals $R = R_1 \cdots R_s, M_1, \dots, M_t$ are pairwise comaximal.
}
\env{proof}{
    Apply \autoref{primes-of-cxy} to \autoref{minimal-decomposition} to obtain $A = R_1 \cap \dots \cap R_s \cap M_1 \cap \dots \cap M_t$ for $R_i$, $M_j$ as in the statement. Observe that $\prod R_i = \bigcap R_i$ since $R_i$ are generated by irreducible polynomials. The ideals $R, M_1, \dots, M_t$ are pairwise comaximal since a maximal ideal is comaximal with any ideal not contained in it. Therefore $A = R \, M_1 \cdots M_t$. The uniqueness follows from uniqueness of minimal decomposition.
\qed}

\env{theorem}{
    \label{ann-is-radical}
    Let $c$ be a two-dimensional finitary integral configuration with a non-trivial annihilator. Then $\Ann(c)$ is a radical ideal. Moreover if $P$ is a prime ideal from the minimal decomposition of $\Ann(c)$ then
    $$P = \gen{x^a y^b - \omega}
    \quad \text{or} \quad P = \gen{x^a - \omega y^b}
    \quad \text{or} \quad P = \gen{x-\omega_x, y-\omega_y}$$
    for $(a,b) \in \N_0^2$ primitive vector and $\omega, \omega_x, \omega_y \in \C$ roots of unity.
}
\env{proof}{
    Denote $A = \sqrt{\Ann(c)}$. Since $c$ has a non-trivial annihilator, $A$ is non-trivial. Let $A = P_1 \cap \dots \cap P_k$ be its minimal decomposition.

    Let $P$ be one of $P_i$. Assume first that $P = \gen{\varphi}$ for an irreducible polynomial $\varphi$. By \autoref{lem-integral-generators} and \autoref{thm-annihilator-from-support} there exist vectors $\vec{u_i} \ne \vec{v_i}$ such that
    $$(X^{\vec{v_1}}-X^{\vec{u_1}}) \cdots (X^{\vec{v_n}}-X^{\vec{u_n}}) \in A.$$
    Since $\varphi$ is an irreducible factor of this polynomial we have $\varphi \mid X^\v - X^\u$ for some $\u \ne \v$. Let $\v-\u = d\w$ for a primitive vector $\w = (a,b)$ and $d > 0$. We can assume $a \geq 0$, otherwise the roles of $\u$ and $\v$ can be exchanged. Observe that in Laurent polynomials
    $$X^\v - X^\u = X^\u(X^{d\w}-1) = X^\u (X^\w - \omega_1) \cdots (X^\w - \omega_d)$$
    where $\omega_1,\dots,\omega_d$ are $d$-th roots of unity. Therefore the irreducible polynomial factors of $X^\v - X^\u$ are, up to a constant multiple, of the form
    \env{align*}{
    	x^a y^b-\omega \ \ \text{(if $b \geq 0$),} \quad \text{or} \quad
    	x^a-\omega y^{-b} \ \ \text{(if $-b > 0$),} \quad \text{or} \quad
    	x, \quad \text{or} \quad y
    }
    for $\omega$ a root of unity. The cases $\varphi = x$ and $\varphi = y$ cannot happen. This classifies the case of principal ideals $P$.
    
    Now assume that $P = \gen{x-\alpha, y-\beta}$ for some $\alpha, \beta \in \C$, without loss of generality let $P = P_1$. Choose $g \in \prod_{i=2}^k (P_i \setminus P_1)$ arbitrarily, then $g(x-\alpha) \in A$ and $g \notin A$. There exists $m \in \N$ such that $g^m(x-\alpha)^m \in \Ann(c)$, but $g^m \notin \Ann(c)$, and in particular $\alpha \ne 0$. In other words, $(x-\alpha)^m$ annihilates the non-zero finitary configuration $c' = g^m c$. By \autoref{lem-prod-of-1d-radicality} also $x-\alpha$ annihilates $c'$, and therefore for every $i,j \in \Z$
    $$c'_{i,j} = c'_{0,j} \alpha^{-i}.$$
    If $\alpha$ is not a root of unity then $c'$ is not finitary, which is a contradiction. A similar argument applies to $\beta$.

    To prove the radicality of $\Ann(c)$, observe that each $P_i$ is generated by line polynomials. Because by \autoref{radical-decomposition-for-cxy} we have $A = P_1 \cdots P_k$, $A$ has a finite set of generators $A = \gen{g_1, \dots, g_k}$ such that each $g_i$ is a product of line polynomials. Then for each $i$ there exists $m \in \N$ such that $g_i^m \in \Ann(c)$, and by \autoref{lem-prod-of-1d-radicality} we have $g_i \in \Ann(c)$. $\Ann(c)$ contains a set of generators of its radical, and therefore it is a radical ideal.
    
\qed}

The proof of the radicality of $\Ann(c)$ relies on the decomposition of two-dimensional radical ideal into a product of primes. Although no analog of such statement is available in higher dimensions, we conjecture that $\Ann(c)$ is radical for higher dimensional finitary configurations as well.

\env{lemma}{
    \label{comaximal-decomposition}
    Let $c$ be a configuration and $A_1, \dots, A_k$, $k \geq 2$ pairwise comaximal ideals such that $\Ann(c) = A_1 \cap \dots \cap A_k$. Then there are uniquely determined configurations $c_1, \dots, c_k$ such that $\Ann(c_i) = A_i$ and $c = c_1 + \dots + c_k$.
}
\env{proof}{
    Note that $\Ann(c) = A_1 \cdots A_k$. We use the following two easy to prove facts from commutative algebra. If $A_i$ are parwise comaximal then:
    \env{enumerate}{
        \item[$(a)$] The ideals $A_1$ and $A_2 \cdots A_k$ are comaximal.
        \item[$(b)$] There exist $f_1, \dots, f_k$ such that $f_i \notin A_i$, $f_i \in \prod_{j \ne i}A_j$ and $f_1 + \dots + f_k = 1$.
    }
    Let $f_i$ be as in $(b)$ and set $c_i = f_ic$. Then $c = c_1 + \dots + c_k$. Let us show $A_1 \subset \Ann(c_1)$:
    $$g \in A_1
        \ \ \Rightarrow\ \ gf_1 \in A_1 \cdots A_k = \Ann(c)
        \ \ \Rightarrow\ \ g \in \Ann(f_1c) = \Ann(c_1).$$
    Next let us show $\Ann(c_1) \subset A_1$. Note that $(1-f_1) = f_2 + 
    \dots + f_k \in A_1$ and compute:
    $$g \in \Ann(c_1)
        \ \ \Rightarrow\ \ gf_1 \in \Ann(c) \subset A_1
        \ \ \Rightarrow\ \ g = gf_1 + g(1-f_1) \in A_1.$$
    For the uniqueness assume $c = c_1' + \dots + c_k'$ such that $c_1 \ne c_1'$ and $\Ann(c_i') = A_i$. By $(a)$ let $f \in A_1$ and $g \in A_2 \cdots A_k$ be such that $f+g=1$. Then
    \env{align*}{
        c_1 - c_1' &= f(c_1 - c_1') + g(c_1 - c_1') \\
        &= f(c_1 - c_1') + g(- c_2 - \dots -c_k + c_2' + \dots + c_k') = 0.
    }
    The argument can be repeated for all $c_i$.
\qed}

\env{note*}{
    If $\Ann(c)$ consisted of Laurent polynomials instead of ordinary polynomials, the statement of \autoref{ann-is-radical} would simplify -- all principal prime ideals in the decomposition would be of the form $\gen{X^\u - \omega}$ for a primitive vector $\u$ (with possibly negative coordinates) and root of unity $\omega$. In the next proof we also deal with the fact that $\Ann(c)$ does not consist of Laurent polynomials, which is done by a technical trick.
}

\envparam[Two-dimensional decomposition theorem]{theorem}{
    \label{thm-2d-decomposition}
    Let $c$ be as in \autoref{ann-is-radical} and $P_1 \cap \dots \cap P_k$ be the minimal decomposition of $\Ann(c)$. Then there exist configurations $c_1, \dots, c_k$ such that $\Ann(c_i) = P_i$ and $c = c_1 + \dots + c_k$.
}
\env{proof}{
    Let $R_1, \dots, R_s$, $M_1, \dots, M_t$ be as in \autoref{radical-decomposition-for-cxy}. By the same corollary, the ideals $R = \prod R_i, M_1, \dots, M_t$ are pairwise comaximal, and by \autoref{comaximal-decomposition} there are configurations $c_R, c_{M_1}, \dots, c_{M_t}$ annihilated by corresponding ideals such that
    $c = c_R + c_{M_1} + \dots + c_{M_t}.$
    
    By \autoref{ann-is-radical}, $R_i = \gen{\varphi_i}$ for some line polynomial $\varphi_i$. These polynomials are in finitely many distinct directions $m$. Define $\phi_1, \dots, \phi_m$ such that each $\phi_j$ is product of all $\varphi_i$ in the same direction. Then, by \autoref{lem-decomposition-with-polynomials}, there are $c_{\phi_1}, \dots, c_{\phi_m}$ annihilated by corresponding polynomials such that
    $c_R = c_{\phi_1} + \dots + c_{\phi_m}.$
    
    Moreover $\Ann(c_{\phi_i}) = \gen{\phi_i}$: if $f \in \Ann(c_{\phi_1})$, then $f\phi_2 \cdots \phi_m \in \Ann(c_R) = R$. The ideal $R$ is one-generated, so $\phi_1 \cdots \phi_m \mid f \phi_2 \cdots \phi_m$ and therefore $f \in \gen{\phi_1}$. Analogously for other $\phi_i$.
    
    For the next step define $S_1 \subset \{1,\dots,s\}$ such that $\phi_1 = \prod_{i \in S_1} \varphi_i$. Since all $\varphi_i$ for $i \in S_1$ have the same direction, by \autoref{ann-is-radical} either they are all of the form $\varphi_i = x^a y^b - \omega_i$ or they are all of the form $\varphi_i = x^a - \omega_i y^b$ for some $a, b \in \N_0$. Assume the first case. Then $\gen{\varphi_i} = R_i$ are pairwise comaximal and by \autoref{comaximal-decomposition} there exist $c_{R_i}$ annihilated by $R_i$ such that $c_{\phi_1} = \sum_{i \in S_1} c_{R_i}$.

    If we have $\varphi_i = x^a - \omega_i y^b$ for $i \in S_1$ we do the following technical trick. Consider the configuration $c'_{\phi_1}$ obtained by mirroring $c_{\phi_1}$ along the horizontal axis. It is easy to verify that $\Ann(c'_{\phi_1}) = \prod_{i \in S_1} \gen{\varphi_i'}$ where $\varphi_i' = x^a y^b - \omega_i$. Proceeding as in the previous case we obtain $c'_{R_i}$ such that $c'_{\phi_1} = \sum_{i \in S_1} c'_{R_i}$ and $\Ann(c'_{R_i}) = \gen{\varphi_i'}$. Let us mirror each $c'_{R_i}$ back along the horizontal axis to obtain $c_{R_i}$. Then $c_{\phi_1} = \sum_{i \in S_1} c_{R_i}$ and $\Ann(c_{R_i}) = R_i$, as desired.
    
    Analogously we can decompose each $c_{\phi_i}$. To finish the proof observe that
    $$c = c_{R_1} + \dots + c_{R_s} + c_{M_1} + \dots + c_{M_t}.$$
\qed}

We say that a two-dimensional configuration is \emph{doubly periodic} if there are two linearly independent vectors in which it is periodic. A configuration which is periodic but not doubly periodic is called \emph{one-periodic}.

\env{corollary}{
    \label{explicit-2d-decomposition}
    Let $c$ be as in \autoref{ann-is-radical}.
    \env{enumerate}{
        \item[$(a)$] There exist a non-negative integer $m$, line polynomials $\phi_1,\dots,\phi_m$ in pairwise distinct directions, a polynomial $\phi := \phi_1 \cdots \phi_m$ and an ideal $H$ which is an intersection of maximal ideals such that $\gen{\phi}$ and $H$ are comaximal and
        $$\Ann(c) = \phi_1 \cdots \phi_m H = \phi H.$$
        Moreover $m$ and $H$ are determined uniquely and $\phi, \phi_1, \dots, \phi_m$ are determined uniquely up to a constant factor and the order.
    
    \item[$(b)$] There exist configurations $c_\phi, c_H, c_1, \dots, c_m$ such that
    $$c = c_1 + \dots + c_m + c_H = c_\phi + c_H$$
    where $\Ann(c_\phi) = \gen\phi$, $\Ann(c_H) = H$ and $\Ann(c_i) = \gen{\phi_i}$.
    Moreover $c_\phi$ and $c_H$ are determined uniquely. Each $c_i$ is one-periodic in the direction of $\phi_i$, and $c_H$ is doubly periodic.
    }
}
\env{proof}{
    Let us continue with the notation from the proof of \autoref{thm-2d-decomposition}.

    $(a)$\ \ Let $H = \bigcap_{i = 1}^t M_i$. Then $\phi, \phi_1, \dots, \phi_m, H$ are as desired.
    
    $(b)$\ \ Let $c_H = c_{M_1} + \dots + c_{M_t}$, $c_\phi = c_R$ and $c_i = c_{\phi_i}$. The fact that $\Ann(c_H) = H$ follows by \autoref{lem:sum-lemma} introduced later and the uniqueness of $c_\phi$ and $c_H$ follows by \autoref{comaximal-decomposition}.
    
    Let $\v$ be a primitive direction of the polynomial $\phi_1$. There is $n \in \N$ such that each irreducible factor of $\phi_1$ divides $X^{n\v} - 1$. Therefore this Laurent polynomial annihilates $c_{\phi_1}$ which means that $c_{\phi_1}$ has period $n \v$. If there was a period $\u$ in any other direction, then $\phi_1 \mid X^\u - 1$, which is impossible. Therefore $c_{\phi_1}$ is one-periodic, and so is any $c_{\phi_i}$.
    
    Denote $M_1 = \gen{x-\omega_x, y-\omega_y}$ and let $n \in \N$ be such that $\omega_x^n = 1$. Then $c_{M_1}$ has a horizontal period $n$ since $x^n-1 \in M_1$. Similarly $c_{M_1}$ has a vertical period. By a similar argument each $c_{M_j}$ is doubly periodic. A finite sum $c_H$ of doubly periodic configurations is also doubly periodic.
\qed}

Let us denote the number $m$ from \autoref{explicit-2d-decomposition} by $\opc(c)$. It is an important characteristic of the configuration which provides information about its periodicity.

\env{corollary}{
    \label{cor-num-of-1per-directions}
    Let $c$ be as in \autoref{ann-is-radical}. Then
    \env{itemize}{
        \item $\opc(c)=0$ if and only if $c$ is doubly periodic,
        \item $\opc(c)=1$ if and only if $c$ is one-periodic,
        \item $\opc(c)\geq 2$ if and only if $c$ is non-periodic.
    }
}

\env{proof}{
	If $\opc(c) = 0$ then $c = c_H$, which is doubly periodic. If $\opc(c) = 1$ then $c = c_1 + c_H$ is a sum of one-periodic and doubly periodic configuration, which is one-periodic. If $\opc(c) \geq 2$ then every annihilating polynomial is divisible by $\phi_1\phi_2$. Therefore $X^\v-1$ cannot be an annihilator for any non-zero vector $\v$ and $c$ is non-periodic.
\qed}

\autoref{explicit-2d-decomposition} and \autoref{cor-num-of-1per-directions} are powerful tools to analyze configurations from the structure of their annihilator ideals. The main improvement over the earlier decomposition theorem is that not only we know that $c$ can be decomposed into a sum of periodic components, but also we can exactly describe the annihilator ideals of each component. Moreover each component is either one- or doubly periodic and the number of one-periodic components (in distinct directions) is unique and determines whether the original configuration is periodic or not.

\env{example}{
    Let us call $D \subset \Z^2$ a \emph{T-shape} if it is of the form
        $$D = \{0,\dots,w\} \times \{h\} \cup \{d\} \times \{0,\dots,h\}$$
    for some $h,w,d \in \N$, $d \leq h$ (\autoref{fig:tshape}). We show that if $P_c(D) \leq \abs D$ for a T-shape $D$, then $c$ is periodic. For a contradiction assume that the inequality holds and $c$ is non-periodic.
    
    We need a fact which is later proved in the next section as \autoref{lem:no-annihilator-high-cplx}: The coefficients of $c$ can be renamed such that if $P_c(D) \leq \abs D$, then there is an annihilator polynomial with $\supp(f) \subset -D$. Without loss of generality assume that the coefficients of $c$ have been renamed and we have such an annihilator $f$.
    
    By \autoref{cor-num-of-1per-directions}, $\opc(c) \geq 2$, and in particular there are two line polynomials $\phi_1$, $\phi_2$ in distinct directions such that $\phi_1 \phi_2$ divides any annihilator polynomial. The convex hull of $\supp(\phi_1 \phi_2)$ is a parallelogram, and therefore the convex hull of $\supp(f)$ has two pairs of parallel sides because $f$ is a polynomial multiple of $\phi_1 \phi_2$. This is, however, impossible since $\supp(f) \subset -D$ and convex hull of any non-collinear subset of points in $-D$ is a triangle.
\qed}

\envparam{figure}{
    \centering
    \includegraphics[scale=0.8]{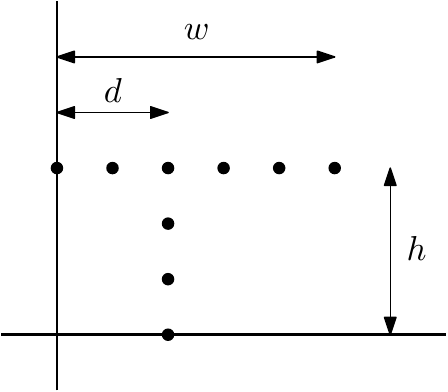}
    \caption{A T-shape with $w = 5$, $h = 3$ and $d = 2$. Convex hull of any non-collinear subset of its points is a triangle.}
    \label{fig:tshape}
}

Knowing a configuration and its annihilator, \autoref{thm-2d-decomposition} gives a decomposition into a sum of configurations and provides their annihilators. We finish the section by giving a complementary claim: given configurations and their annihilators, we can describe the annihilator of their sum.

\env{lemma}{
    \label{lem:sum-lemma}
    Let $c_1, c_2$ be configurations such that $\Ann(c_1)$ and $\Ann(c_2)$ are non-trivial radical ideals. Let $P_1, \dots, P_k$, $Q_1, \dots, Q_\ell$ be prime ideals such that
        $$\Ann(c_1) = \bigcap_{i=1}^k P_i \qquad \text{and} \qquad \Ann(c_2) = \bigcap_{j=1}^\ell Q_j$$
    are minimal decompositions. If $P_i \ne Q_j$ for all admissible $i,j$, then $\Ann(c_1+c_2) = \Ann(c_1) \cap \Ann(c_2)$.
}
\env{proof}{
    Denote $c = c_1 + c_2$, clearly $\Ann(c) \supset \Ann(c_1) \cap \Ann(c_2)$. To prove the other inclusion, for the contrary suppose there exists $f \in \Ann(c)$ such that $f \notin \Ann(c_1) \cap \Ann(c_2)$. Then $f$ does not belong to at least one of the prime ideals. Without loss of generality assume $f \notin P_1$ and $P_1$ is minimal such ideal with respect to inclusion. In particular, we have $Q_j \nsubseteq P_1$ for every $j$.
    
    Now choose any $g \in \prod_{j=1}^\ell \left( Q_j \setminus P_1 \right)$, then we have $g \in \Ann(c_2) \setminus P_1$. Consider the polynomial $fg$. Since $f$ annihilates $c$ and $g$ annihilates $c_2$, we have that $fg$ annihilates $c - c_2 = c_1$. But $fg \notin P_1$, which is in contradiction with $\Ann(c_1) \subset P_1$.
\qed}

\env{corollary}{
	\label{cor:sum-lemma-ord}
    Let $c_1, c_2$ be two-dimensional finitary integral configurations having a non-trivial annihilator and $k = \opc(c_1)$, $\ell = \opc(c_2)$ such that
        $$\Ann(c_1) = \phi_1 \cdots \phi_k H_1 \qquad \text{and} \qquad \Ann(c_2) = \psi_1 \cdots \psi_\ell H_2$$
    where $\phi_i, \psi_j$ are line polynomials and $H_1, H_2$ intersections of maximal ideals as in \autoref{explicit-2d-decomposition}. If $\phi_i$ and $\psi_j$ have pairwise distinct directions, then $\opc(c_1+c_2) = k+\ell$ and there exists $H$ an intersection of maximal ideals such that
    $$\Ann(c_1+c_2) = \phi_1 \cdots \phi_k \psi_1 \cdots \psi_\ell H.$$
\qed}

\env{example}{
    \label{ex:sum-of-oneperiodic} Let us show that if $c_1$ and $c_2$ are two-dimensional finitary one-periodic configurations in distinct directions, then their sum is non-periodic.
    
    By \autoref{cor-num-of-1per-directions} we have $\opc(c_1) = \opc(c_2) = 1$, and therefore by \autoref{explicit-2d-decomposition} there are $\phi, \psi$ line polynomials such that $\Ann(c_1) = \phi H_1$ and $\Ann(c_2) = \psi H_2$ for some $H_1, H_2$ intersections of maximal ideals. Moreover $\phi$ and $\psi$ have the same direction as is the unique direction of periodicity of $c_1$ and $c_2$ respectively. Therefore, by the previous lemma, $\opc(c_1+c_2) = 2$ and therefore $c_1+c_2$ is non-periodic by \autoref{cor-num-of-1per-directions}.
\qed}

\section{Approaching Nivat's Conjecture}
\label{sec-results}


In this section we apply the facts we learned in previous sections about annihilating polynomials and link them to the complexity of a configuration.

When going from a symbolic configuration to formal power series, we have to choose numerical representations of the symbols. We begin by showing that there is a particularly suitable choice, and we call such configurations \emph{normalized}. Next, in order to attack Nivat's conjecture, we define a class of configurations called \emph{counterexample candidates}. As the name suggests, these are potential counterexamples to the conjecture, and our goal is to prove that such configurations have high complexity.

To handle the complexity we need a suitable tool. We introduce \emph{lines of blocks}, which are just sets of blocks $m \times n$ located on a common line in the configuration. We prove two complementary lemmas -- the first one states that there are many disjoint lines of blocks, while the other gives a lower bound on the number of distinct blocks on a line. These combined result in a lower bound on the overall complexity.

Our main result is that if $c$ is non-periodic then the condition $P_c(m,n) > mn$ is true for all but finitely many pairs $m,n$. In the proof we consider three different ranges of $m$ and $n$:
\smallskip

\noindent
{\bf Very thin blocks.} If $m$ or $n$ is so small that the support of no annihilating polynomial fits in the $m \times n$ rectangle, then by a variation of \autoref{lem-low-cplx-has-annihilator} the configuration has complexity $P_c(m,n) > mn$.
\smallskip

\noindent
{\bf Thin blocks.} Consider fixed $n$, large enough so that the support of some annihilator fits inside a strip of height $n$. We show that there exists $m_0$ such that for all $m>m_0$ we have  $P_c(m,n) > mn$. Analogously for a fixed $m$.
\smallskip

\noindent
{\bf Fat blocks.} We prove that there are constants $m_0$ and $n_0$ such that for $m>m_0$ and $n>n_0$ we have $P_c(m,n) > mn$.
\smallskip

These three ranges cover all but finitely many dimensions $m \times n$. Interestingly, a common approach works for all configurations except for the case of fat blocks when $c$ is a sum of horizontally and vertically one-periodic configuration. This case requires a more involved combinatorial analysis which is carried out separately in \autoref{sec-rectilinear-case}.

\subsection*{Normalized Configurations}

There is a particularly suitable choice when representing a symbolic configuration as a formal power series. For a configuration $c$ consider Laurent polynomials $f$ such that $fc$ is a constant configuration. We say that $c$ is \emph{normalized} if all such $f$ are annihilators, i.e. the constant in the result of $fc$ is zero. Let us denote by $\mathbbm1$ the constant one configuration.

\env{lemma}{
    \label{lem:normalized}
    Let $c$ be a finitary configuration. Then there exists $a,b \in \C, a \ne 0$ such that $ac + b \mathbbm1$ is normalized. Moreover if $c$ is integral then $a,b \in \Z$.
}
\env{proof}{
    Let $f,g$ be Laurent polynomials such that $fc, gc$ are constant configurations. Denote by $\kappa(f)$ the number such that $fc = \kappa(f)\mathbbm1$ and by $\sigma(f)$ the sum of the coefficients of $f$. Then
    \env{gather*}{
        \sigma(f)\kappa(g)\mathbbm1 = fgc = gfc = \sigma(g)\kappa(f)\mathbbm1 \\
        \Rightarrow \quad g\big(\sigma(f)c - \kappa(f)\mathbbm1\big) = \sigma(f)\kappa(g)\mathbbm1 - \kappa(f)\sigma(g)\mathbbm1 = 0.
    }
    
    If there is $f$ such that $\sigma(f) \ne 0$ we can choose $a = \sigma(f), b = -\kappa(f)$ and we are done. Let us assume that for all $f$ we have $\sigma(f) = 0$, we will show that then $c$ is already normalized and therefore we can choose $a = 1, b = 0$.
    
    For even $k$ let $C_k$ denote the hypercube $[-\frac{k}{2},\frac{k}{2})^d \subset \Z^d$ of side $k$ centered around the origin. Choose even $n \in \N$ such that $\supp(f) \subset C_n$ and consider arbitrary even integer $N > n$. Let us count the sum of coefficients of $fc$ inside of $C_N$.

\envparam{figure}{
    \centering
    \includegraphics[scale=0.9]{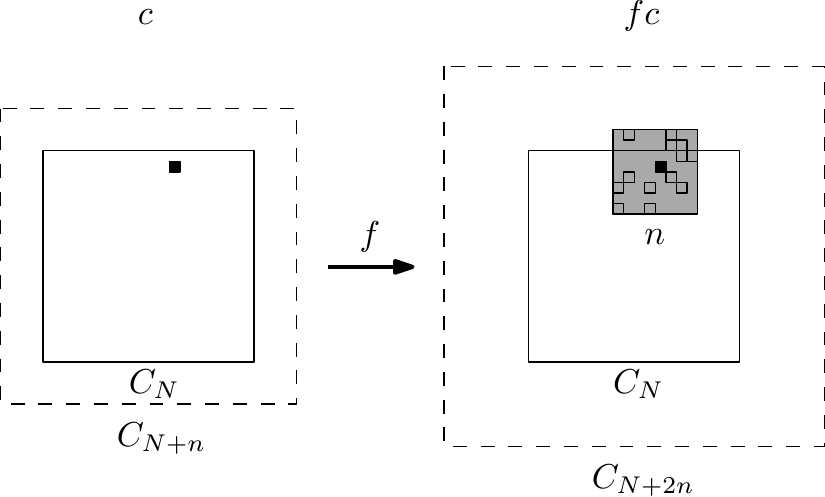}
    \caption{Proof of \autoref{lem:normalized}: Counting sum of coefficients of $fc$ inside of $C_N$.}
    \label{fig:normalized-proof}
}    
    
    Since $fc$ is a constant configuration the sum is surely $\kappa(f)N^d$. On the other hand, the coefficients of $fc$ in $C_N$ depend only on the coefficients of $c$ in $C_{N+n}$. Each such coefficient $c_\v$ contributes to the sum by $\sigma(f)c_\v$, but we overcount in the region $C_{N+2n} \setminus C_N$ of $fc$, see \autoref{fig:normalized-proof}. This region is of size proportional to $N^{d-1}$ and because $c$ is finitary, the contribution to each position is bounded. Therefore
    \env{align*}{
        \kappa(f)N^d = \sum_{\v \in C_{N+n}} \sigma(f)c_v + O(N^{d-1}) = O(N^{d-1}).
    }
    Taking the limit $N \rightarrow \infty$ shows that $\kappa(f) = 0$. Therefore $f$ is an annihilator and $c$ is normalized.
    
    For the "moreover" part we argue as in the proof of \autoref{lem-integral-generators}. Let $f = \sum a_i X^{\vec{u_i}}$, then
    \env{align*}{
        fc = a_0\mathbbm1 \quad \Leftrightarrow \quad (-\overline{a_0}, \overline{a_1}, \dots, \overline{a_m}) \perp (1, c_{\v-\vec{u_1}}, \dots, c_{\v-\vec{u_m}})
    }
    for all $\v \in \Z$. Thus all $f$ form a vector space over $\C$ which has integral generators if $c$ is integral. Therefore if there is $f$ with $\sigma(f) \ne 0$, then there is also integral $f'$ with $\sigma(f') \ne 0$. In that case necessarily $\sigma(f'), \kappa(f') \in \Z$.
\qed}

\env{corollary}{
	Either $c$ is normalized, in which case $c + \kappa\mathbbm1$ is normalized for all choices of $\kappa \in \C$, or there is unique $\kappa \in \C$ such that $c + \kappa\mathbbm1$ is normalized.
}
\env{proof}{
    Follows from the proof of \autoref{lem:normalized} by choosing $\kappa = b/a$.
\qed}

Note that the case when $\sigma(f) = 0$ for all $f$ in the proof of the previous lemma can be handled easily for two-dimensional integral configurations. If the sum of coefficients of $f$ is zero and $fc$ is a constant configuration, then $f^2c = 0$. We proved that the ideal of annihilators is radical, so we can conclude $fc = 0$.

\medskip

To link polynomials and complexity we use a variation of \autoref{lem-low-cplx-has-annihilator}. Recall that for a finite shape $D \subset \Z^d$ we denote by $c_{\v+D}$ the pattern of shape $D$ extracted from the position $\v \in \Z^d$. Formally we defined it as a function
\env{align*}{
    c_{\v+D}: D &\rightarrow \C \\
    \vec{d_i} &\mapsto c_{\v+\vec{d_i}},
}
and therefore it makes sense to talk about linear independence of patterns (over $\C$). If we denote $D = \{\vec{d_1}, \dots, \vec{d_n}\}$, then this is the same as if we considered $c_{\v+D}$ to be the vector $(c_{\v+\vec{d_1}}, \dots, c_{\v+\vec{d_n}}) \in \C^n$.

Let us say that a Laurent polynomial $f$ \emph{fits in} $S \subset \Z^d$ if a translate of $-\supp(f)$ is a subset of $S$. Here $S$ can also be infinite, and usually will be a convex subset of $\Z^d$.

\env{lemma}{
    \label{lem:no-annihilator-high-cplx}
    Let $c$ be a configuration and $D \subset \Z^d$ a finite shape. Assume there is no annihilating Laurent polynomial $f$ which fits in $D$. Then there are $\abs D$ linearly independent patterns $c_{\v+D}$. Moreover if $c$ is normalized then $P_c(D) > \abs D$.
}
\env{proof}{
    Denote $D$ as above and for contradiction assume the vectors $(c_{\v+\vec{d_1}}, \dots, c_{\v+\vec{d_n}}) \in \C^n$ span a space of dimension at most $n-1$. Then there exists a common orthogonal vector $(\overline{a_1},\dots,\overline{a_n})$ and $f(X) = a_1 X^{-\vec{d_1}} + \dots + a_n X^{-\vec{d_n}}$ is an annihilating polynomial fitting in $D$.
    
    For the second part for contradiction suppose $P_c(D) \leq n$, then the vectors $(1, c_{\v+\vec{d_1}}, \dots, c_{\v+\vec{d_n}}) \in \C^{n+1}$ span a space of dimension at most $n$. Let $(\overline{a_0},\overline{a_1},\dots,\overline{a_n})$ be their common orthogonal vector. Then $f$ defined as previously has the property $fc = -a_0\mathbbm1$. If $c$ is normalized then $f$ is an annihilator.
\qed}

\subsection*{Counterexample Candidates}

We approach Nivat's conjecture by examining a potential counterexample to it. Let us recall the conjecture, in the contrapositive direction:

\envparam[Nivat's conjecture]{conjecture*}{
    Let $c$ be a non-periodic two-dimensional configuration. Then for all positive integers $m,n$ we have $P_c(m,n) > mn$.
}

If $c$ is a counterexample, then it is surely a non-periodic two-dimensional configuration. It is finitary, since otherwise its complexity is not bounded. It also has to have an annihilator -- otherwise by \autoref{lem-low-cplx-has-annihilator} for all $m, n$ we have $P_c(m,n) > mn$. Moreover, without loss of generality, we can assume that $c$ is integral. Let us make a formal definition:

\env{definition}{ \label{def:counterexample}
A configuration is a \emph{counterexample candidate} if it is two-dimensional, non-periodic, finitary and integral configuration with an annihilator.
}

Our goal is to show that any counterexample candidate $c$ has a high complexity. In the proofs which follow we will frequently use the annihilator structure characterization from \autoref{explicit-2d-decomposition}. Let us therefore define polynomials $\phi, \phi_1, \dots, \phi_{\opc(c)}$ and an ideal $H$ such that
\env{align*}{
    \Ann(c) = \phi H = \phi_1 \cdots \phi_{\opc(c)} H
}
as in the statement of \autoref{explicit-2d-decomposition}. Note that since $c$ is non-periodic we have $\opc(c) \geq 2$.

For a non-zero Laurent polynomial $f$ let us define the \emph{bounding box} of $f$ to be the vector $\bbox(f) = (m,n)$ with $m,n$ smallest integers such that $f$ fits in a block $(m+1) \times (n+1)$. Equivalently,
\env{align*}{
    \bbox(f) = (\max A - \min A,\ \max B - \min B)
}
where $A = \makeset{a}{(a,b) \in \supp(f)}$ and $B = \makeset{b}{(a,b) \in \supp(f)}$. Let us furthermore extend the definition to vectors: if $\v = (v_1,v_2)$ then define $\bbox(\v) = (\abs{v_1}, \abs{v_2})$.

\env{example}{
    For example, $\bbox(xy^{-1} + x^2y - 3x^3) = (2,2)$ and $\bbox(X^\u-X^\v) = \bbox(\u-\v)$. If we plot the support of a polynomial as points in the plane, the bounding box are dimensions of the smallest rectangle which covers all of them, see \autoref{fig:ex-bounding-box}. Note however that a polynomial $f$ never fits in $\bbox(f)$.
}

\envparam{figure}{
    \centering
    \includegraphics[scale=0.8]{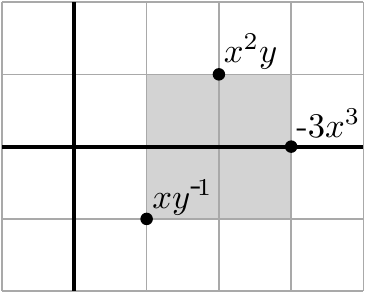}
    \caption{The bounding box of the polynomial $xy^{-1} + x^2y - 3x^3$ is $(2,2)$.}
    \label{fig:ex-bounding-box}
}

With the framework that we just defined we get almost for free that counterexample candidates have high complexity for very thin rectangles:

\envparam[Very thin blocks]{lemma}{
    \label{lem:very-thin-blocks}
    Let $c$ be a counterexample candidate and $(m_\phi,n_\phi) = \bbox(\phi)$. If $M,N$ are positive integers such that $M \leq m_\phi$ or $N \leq n_\phi$ then $P_c(M,N) > MN$.
}
\env{proof}{
    By \autoref{lem:normalized} there exist $a,b \in \Z, a \ne 0,$ such that $c' = ac+b\mathbbm1$ is a finitary integral configuration which is normalized. Clearly $P_c(M,N) = P_{c'}(M,N)$. Let $\Ann(c') = \phi'H'$. Since $\Ann(ac) = \Ann(c)$ and $\opc(b\mathbbm1) = 0$, by \autoref{cor:sum-lemma-ord} we have $\phi' = \phi$.
    
    Thus every annihilator of $c'$ is a multiple of $\phi$ and therefore it cannot fit in an $M \times N$ rectangle. By \autoref{lem:no-annihilator-high-cplx} we have $P_{c'}(M,N) > MN$ which concludes the proof.
\qed}

\subsection*{Disjoint Lines of Blocks}

For a finite shape $D \subset \Z^2$ let us define a \emph{line of $D$-patterns in direction $\v \in \Z^2$}, $\v \ne 0$ to be a set of the form
\env{align*}{
    \L = \makesetbig{c_{\u+k\v+D}}{k \in \Z}
}
for some vector $\u \in \Z^2$. Let $Lines_\v(D)$ be the set of all lines in the same direction, i.e.
\env{align*}{
    Lines_\v(D) = \makesetbig{\makeset{c_{\u+k\v+D}}{k \in \Z}}{\u \in \Z^2}.
}
Note that $Lines_\v(D)$ is a family of sets. In our usual setup the vector $\v$ will be primitive and as the shape $D$ we will consider rectangular blocks $M \times N$. In that case we talk about \emph{lines of $M \times N$ blocks in direction $\v$} and denote more conveniently by $Lines_\v(M,N)$. \autoref{fig:lines} illustrates this definition.

\envparam{figure}{
    \centering
    \includegraphics[scale=1.0]{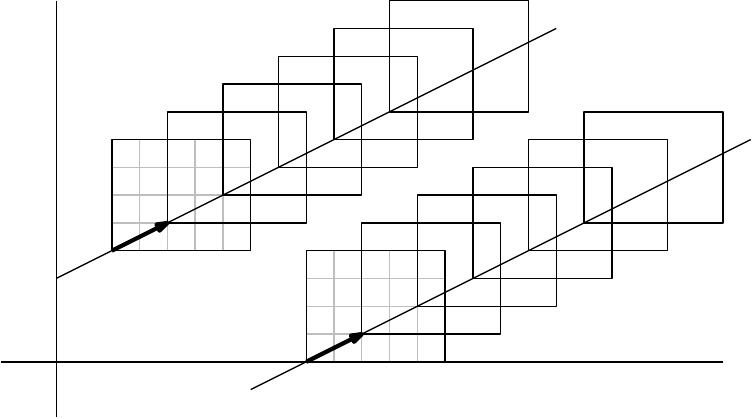}
    \caption{Two lines of blocks $5 \times 4$ in direction $(2,1)$. They are elements of $Lines_{(2,1)}(5,4)$.}
    \label{fig:lines}
}

Our strategy is to prove two complementary lemmas. The first one gives a lower bound on the number of pairwise disjoint sets in $Lines_\v(M,N)$ for a suitable choice of $\v, M, N$. The second one gives a lower bound for the number of blocks in any $\L \in Lines_\v(M,N)$. Combined, they give a lower bound on the complexity of the configuration.

We make use of the structure of the annihilator ideal $\Ann(c) = \phi H$. When talking about \emph{minimal} polynomials, we mean minimal with respect to polynomial division. In polynomials in one variable, all ideals have (up to a constant factor) unique minimal polynomial which generates the ideal. In our case the situation can be more complicated.

Clearly, minimal polynomials of $\Ann(c)$ are of the form $\phi h$ where $h$ is a minimal polynomial of $H$. Moreover, in that case $\Ann(hc) = \gen \phi$. Note that we cannot take any polynomial from $H$ in the place of $h$ -- for example, $\phi h \in H$ but $\Ann(\phi h c) = \Ann(0) = \C[x,y]$.

We claim that $H$ contains a line polynomial in arbitrary non-zero direction $\v \in \Z^2$ which is minimal. If $H = \C[X]$ this is trivially true. Otherwise let $Z_i \in \C^2$ be the roots of $H$. Then for a suitable $\u \in \Z^2$, $X^u\prod_i (X^\v - Z_i^\v) \in H$ is a line polynomial in the direction $\v$. It suffices to choose a minimal polynomial from $H$ which divides it.


\env{lemma}{
    \label{lem:disjoint-lines-oneperiodic}
    Let $f$ be a line Laurent polynomial and $\v$ a primitive vector in the direction of $f$. Let $c$ be a configuration such that $\Ann(c) = \gen f$. Denote $(m_f,n_f) = \bbox(f)$, $(m,n) = \bbox(\v)$ and let $M > m_f, N > n_f$ be positive integers. Then $Lines_\v(M,N)$ contains at least $(M-m_f)n + m(N-n_f)$ pairwise disjoint sets.
}
\env{proof}{
    Without loss of generality assume $\v = (m,n)$, otherwise a mirrored or rotated configuration can be considered. There is an integer $d \in \N$ such that $(m_f,n_f) = (dm, dn) = d\v$. Denote $M' = M - m_f, N' = N - n_f$ and define
    \env{align*}{
        D =  \makesetbig{(M',0) + a(-M',N') + b(m_f,n_f)}{a, b \in [0,1)} \cap \Z^2.
    }
    The shape $D$ is contained in an $M \times N$ block and $\abs D = M' n_f + m_f N'$, see \autoref{fig:disjoint-lines}. Moreover no multiple of $f$ fits in $D$, thus by \autoref{lem:no-annihilator-high-cplx} there are at least $M'n_f+m_fN' = d(M'n+mN')$ linearly independent patterns $c_{\v+D}$.
    
    Let $\L$ be a line of patterns from $Lines_\v(D)$. Then $f$ gives a linear recurrence relation of degree $d$ on the elements of $\L$. Therefore the vector space generated by the elements of $\L$ has dimension at most $d$. In particular, each line contains at most $d$ of the $\abs D$ linearly independent patterns $c_{\v+D}$. It follows that there are at least $M'n+mN'$ distinct lines in $Lines_\v(D)$.
    
    We claim that if two lines are distinct then they are disjoint. Indeed, if a line contains a particular $D$-pattern, then $f$ uniquely determines the next and the previous pattern on the line. Therefore the lines either contain exactly the same patterns or they are disjoint.
    
    We proved that $Lines_\v(D)$ contains at least $M'n+mN'$ pairwise disjoint lines, therefore also $Lines_\v(M,N)$ does.
\qed}

\envparam{figure}{
    \centering
    \includegraphics[scale=1.0]{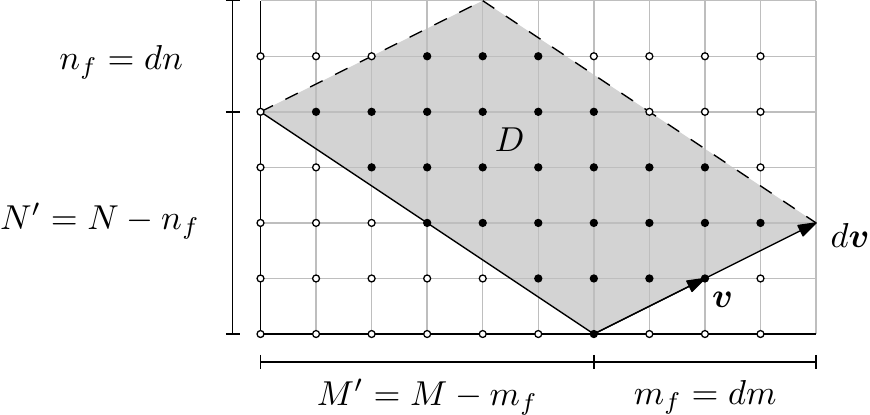}
    \caption{The shape $D$ in \autoref{lem:disjoint-lines-oneperiodic}. The marked points are elements of the $M \times N$ block, the filled ones belong to $D$.}
    \label{fig:disjoint-lines}
}

\env{corollary}{
    \label{cor:disjoint-columns}
    Let $c$ be a vertically one-periodic two-dimensional finitary configuration and $f \in \C[y]$ minimal vertical polynomial which annihilates it. Let $n_f$ be the degree of $f$. Then for any $M > 0, N > n_f$ the family $Lines_{(0,1)}(M,N)$ contains at least $M$ disjoint sets.
}
\env{proof}{
    Let $h \in H$ be such that $f = \phi h$, clearly $h \in \C[y]$. Denote $n_h$ the degree of $h$. Then $\Ann(hc) = \gen{\phi}$ and by \autoref{lem:disjoint-lines-oneperiodic} the set $Lines_{(0,1)}(M,N-n_h)$ in $hc$ contains at least $M$ disjoint columns of blocks. An $M \times N$ block in $c$ determines an $M \times (N-n_h)$ block in $hc$. Therefore also $Lines_{(0,1)}(M,N)$ contains at least $M$ disjoint columns of blocks.
\qed}

\env{lemma}{
    \label{lem:disjoint-lines}
    Let $c$ be a counterexample candidate, $f \in \Ann(c)$ be minimal and $\v$ be a primitive vector in the direction of $\phi_1$. Denote $(m_f,n_f) = \bbox(f)$, $(m,n) = \bbox(\v)$ and let $M > m_f$, $N > n_f$ be integers. Then $Lines_\v(M,N)$ contains at least $(M-m_f)n + m(N-n_f)$ disjoint sets.
}
\env{proof}{
    Let $c' = (f / \phi_1)c$, then $c'$ is a one-periodic configuration with $\Ann(c') = \gen{\phi_1}$. Denote $(m_1, n_1) = \bbox(\phi_1)$, then by \autoref{lem:disjoint-lines-oneperiodic}, $Lines_{\v}(M-m_f+m_1,N-n_f+n_1)$ in $c'$ contains at least $(M-m_f)n+m(N-n_f)$ disjoint elements. An $M \times N$ block in $c$ when multiplied by $f/\phi_1$ determines an $(M-m_f+m_1)\times(M-n_f+n_1)$ block in $c'$. Therefore the lower bound applies also for $Lines_\v(M,N)$ in $c$.
\qed}

\subsection*{Non-periodic Stripes}

Define a \emph{stripe} to be a set of integer points between two parallel lines, i.e. a set of the form
\env{align*}{
    \makesetbig{\vec{w} + a\u + b\v}{a \in [0,1), b \in \R} \cap \Z^2,
}
where $\u, \v, \vec w \in \Z^2$ are arbitrary, $\v \ne 0$. The vector $\vec w$ specifies the position of the stripe, $\u$ determines its width and the stripe extends infinitely along $\v$. Let us call 
the vector space $\gen \v \subset \Q^2$ the \emph{direction} of the stripe.

\env{lemma}{
    \label{lem:nonperiodic-stripe}
    Let $c$ be a counterexample candidate and $\v \in \Z^2$ a non-zero vector. Let $S$ be an infinite stripe in the direction of $\v$ of maximal width such that $\phi$ does not fit in. Then $c$ restricted to the stripe $S$ is non-periodic in the direction of $\v$.
}
\env{proof}{
    Since $\opc(c) \geq 2$ there are at least two line polynomial factors of $\phi$ in different directions. Without loss of generality assume that $\v$ is distinct from the direction of $\phi_1$.
    
    Let $h \in H$ be a minimal line polynomial in the direction of $\v$. Then $f = \phi h$ is a minimal polynomial from $\Ann(c)$. Consider $c' = (f/\phi_1)c$. It is a one-periodic configuration in the direction of $\phi_1$. Let $S'$ be a narrower stripe in $c'$ determined from $S$ in $c$ by the multiplication by $f/\phi_1$. $S'$ is of maximal width such that $\phi_1$ does not fit in.
    
    For a contradiction assume that $c$ restricted to $S$ is periodic in the direction of $\v$, then also $c'$ restricted to $S'$ is. Moreover $S'$ determines the whole configuration $c'$ -- the annihilator $\phi_1$ gives a linear recurrence relation on the coefficients of $c'$ lying on lines in the direction of $\phi_1$, and $S'$ is wide enough so that every coefficient is determined. Therefore $c'$ is periodic also in the direction of $\v$, which is in contradiction with one-periodicity of $c'$.
\qed}

\env{lemma}{
    \label{lem:complex-lines}
    Let $c$ be a counterexample candidate and $\v \in \Z^2$ a non-zero vector. Denote $(m_\phi,n_\phi) = \bbox(\phi)$, $(m,n) = \bbox(\v)$ and let $M>m_\phi$, $N>n_\phi$ be integers. Let $\L \in Lines_\v(M,N)$ be arbitrary.
    \envparam[(a)]{enumerate}{
        \item If $\v$ is neither horizontal nor vertical, then
            \env{align*}{
                \abs \L \geq \min\left\{\frac{M-m_\phi+1}{m}, \frac{N-n_\phi+1}{n} \right\}.
            }
        \item Assume $\v$ is not horizontal. If $M \geq (N+n_\phi)\frac{m}{n} + m_\phi$ then
            \env{align*}{
                \abs \L \geq \frac{N+1}{n}.
            }
    }
}
\env{proof}{
    Without loss of generality assume $\v = (m,n)$, the other cases are mirrored or rotated. Also assume that there is a block in $\L$ with $(0,0)$ as its bottom left corner. The proof is illustrated in \autoref{fig:complex-lines}.
    
    \smallskip\noindent
    (a)\ \ Consider the stripe
    \env{align*}{
        S_1 = \makesetbig{(0,n_\phi)+ a(m_\phi,-n_\phi) + b\v}{a \in [0,1), b \in \R} \cap \Z^2.
    }
    Since $(m_\phi,n_\phi)$ is the bounding box of $\phi$, the stripe $S$ from \autoref{lem:nonperiodic-stripe} fits in $S_1$. Therefore $S_1$ is non-periodic in the direction of $\v$, and in particular there exists a "fiber" $f = \makeset{\u+k\v}{k \in \Z}$ inside of the stripe on which $c$ spells a non-periodic sequence.
            
    Each block from $\L$ contains the same number of consecutive points from a fixed fiber in $S_1$, let $p(f)$ be this number for $f$. Clearly, one of the two fibers on the boundaries of $S_1$ lower bounds this quantity. Therefore, by computing the number of points on the boundary fibers,
    \env{align*}{
        p(f) \geq \min \left\{ \Big\lfloor \frac{M-m_\phi}{m} \Big\rfloor, \Big\lfloor \frac{N-n_\phi}{n} \Big\rfloor \right\}.
    }
    
    Now by Morse-Hedlund theorem there are at least $p(f) + 1$ distinct blocks in $\L$. The proof is finished by verifying that $\lfloor p/q \rfloor + 1 \geq (p+1)/q$ for $p,q \in \N$.

    \smallskip\noindent
    (b)\ \ Consider the stripe
    \env{align*}{
        S_2 = \makesetbig{a(m_\phi,-n_\phi) + b\v}{a \in [0,1), b \in \R} \cap \Z^2.
    }
    As in the part (a), it contains a non-periodic fiber. Moreover, if the condition on $M$ is satisfied, then the boundary of $S_2$ intersects every block in $\L$ on the top edge. Therefore $\lfloor N/n \rfloor$ lower bounds the number of points from any fiber of $S_2$ contained in a block in $\L$. The rest follows as in (a).
\qed}

\envparam{figure}{
    \centering
    \includegraphics[scale=0.8]{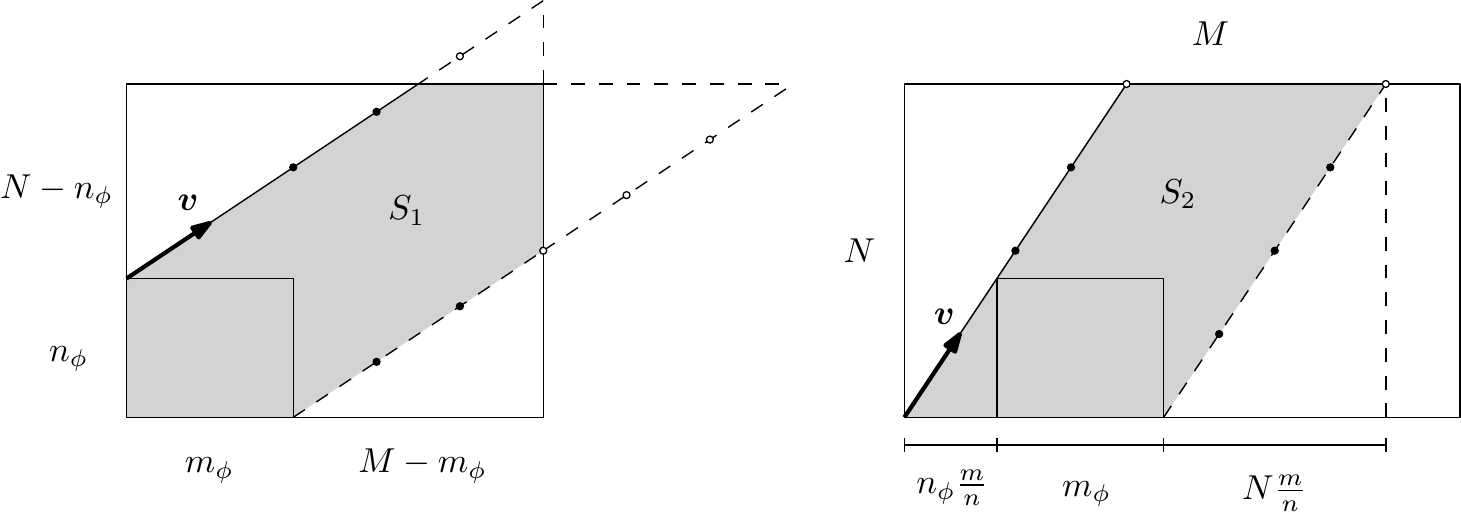}
    \caption{The stripes $S_1$ and $S_2$ from the proof of \autoref{lem:complex-lines}.}
    \label{fig:complex-lines}
}

\subsection*{The Main Result}

Let us combine the above lemmas to get a lower bound on the complexity of a counterexample candidate.

\envparam[Thin blocks]{lemma}{
    \label{lem:thin-blocks}
    Let $c$ be a counterexample candidate and $(m_\phi,n_\phi) = \bbox(\phi)$. Fix an integer $N > n_\phi$. Then there exists $M_0$ such that if $M > M_0$ then $P_c(M,N) > MN$.
}
\env{proof}{
    Since $\opc(c) \geq 2$ we can without loss of generality assume that the direction of $\phi_1$ is not horizontal. Let $\v$ be a primitive vector in that direction and denote $(m,n) = \bbox(\v)$.
    
    Let $h \in H$ be a horizontal line polynomial and let $f = \phi h$, $(m_f, n_f) = \bbox(f)$. Clearly $n_f = n_\phi$. Assume $M \geq (N+n_\phi)\frac{m}{n} + m_\phi$. Then by \autoref{lem:disjoint-lines} and \autoref{lem:complex-lines}(b) for $M > m_f, N > n_f$ we have
    \env{align*}{
         P_c(M,N)
	         &= \left|\, \bigcup Lines_\v(M,N) \,\right| \\
             &\geq \big( (M-m_f)n + m(N-n_f) \big) \frac{N+1}{n} \\
             &\geq (M-m_f)(N+1) = MN + M - m_f(N+1).
    }
    The proof is finished by choosing $M_0 = \max \big\{ m_f(N+1),\ (N+n_\phi)\frac{m}{n} + m_\phi \big\}$.
\qed}

\envparam[Fat blocks I]{lemma}{
    \label{lem:fat-blocks}
    Let $c$ be a counterexample candidate and let $\v$ be the direction of $\phi_1$. If $\v$ is neither horizontal nor vertical, then there exist positive integers $M_0,N_0$ such that for $M > M_0$ and $N > N_0$ holds $P_c(M,N) > MN$.
}
\env{proof}{
    Let $f \in \Ann(c)$ be minimal and denote $(m,n) = \bbox(\v)$, $(m_\phi,n_\phi) = \bbox(\phi)$, $(m_f,n_f) = \bbox(f)$. Assume $M > m_f$, $N > n_f$ and let $\alpha = \frac{m}{n}$. We consider three ranges of $M$. The proof is illustrated in \autoref{fig:fat-blocks}.
    
    \medskip\noindent
    (a)\ \ Assume $(N+n_\phi)\alpha + m_\phi \leq M$. This condition is equivalent to the one in \autoref{lem:complex-lines}(b), therefore by combining with \autoref{lem:disjoint-lines}
    \env{align*}{
         P_c(M,N) &\geq \big( (M-m_f)n + m(N-n_f) \big) \frac{N+1}{n} \\
             &= (M-m_f)(N+1) + (N-n_f)(N+1)\frac{m}{n} \\
             &= MN + M + \Theta(N^2).
    }
    Therefore there exist an integer $N_1$ such that for $N > N_1$ the complexity is at least $MN$.
    
    \medskip\noindent
    (b)\ \ Assume $(N-n_\phi) \alpha - m_\phi < M < (N+n_\phi)\alpha + m_\phi$. Then $M = \Theta(N)$. Now combine \autoref{lem:disjoint-lines} and \autoref{lem:complex-lines}(a):
    \env{align*}{
         P_c(M,N) &> \big( (M-m_f)n + m(N-n_f) \big) \min \left\{ \frac{M-m_\phi}{m}, \frac{N-n_\phi}{n} \right\} \\
          &\geq \big( (M-m_f)n + m(N-n_f) \big) \min \left\{ \frac{M-m_f}{m}, \frac{N-n_f}{n} \right\} \\
         &= (M-m_f)(N-n_f) + \min \left\{ (M-m_f)^2 \frac{n}{m}, (N-n_f)^2 \frac{m}{n} \right\} \\
         &= (M-m_f)(N-n_f) + \Theta(N^2) \\
         &= MN + \Theta(N^2).
    }
    Therefore there is an integer $N_2$ such that for $N > N_2$ the complexity exceeds $MN$.
    
    \medskip\noindent
    (c)\ \ Assume $M \leq (N-n_\phi) \alpha - m_\phi$. This is equivalent to the condition in \autoref{lem:complex-lines}(b) when the roles of horizontal and vertical direction are exchanged. Therefore, similarly as in (a), there exists $M_0$ such that for $M > M_0$ the complexity is at least $MN$. The whole proof is finished by choosing $N_0 = \max\{N_1, N_2\}$.
\qed}

\envparam{figure}{
    \centering
    \includegraphics[scale=0.8]{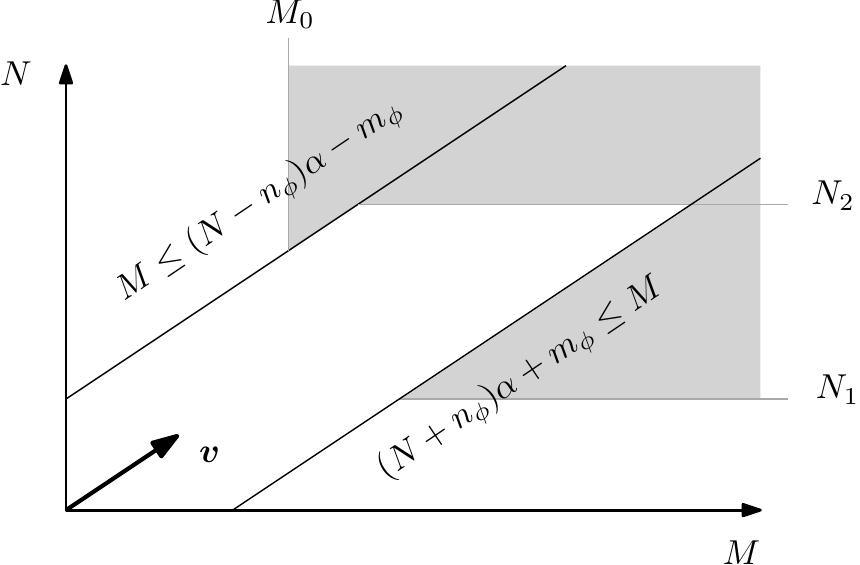}
    \caption{Three different ranges for $M$ from the proof of \autoref{lem:fat-blocks}.}
    \label{fig:fat-blocks}
}

Now we are just a step away from our main theorem. Suppose we knew that \autoref{lem:fat-blocks} holds also when there are only horizontal and vertical $\phi_i$ components:

\envparam[Fat blocks II]{lemma}{
    \label{lem:rectilinear-fat-blocks}
    Let $c$ be a counterexample candidate, $\opc(c) = 2$ and the directions of $\phi_1, \phi_2$ are horizontal and vertical, respectively. Then there exist positive integers $M_0,N_0$ such that for $M > M_0$ and $N > N_0$ holds $P_c(M,N) > MN$.
}

This is exactly the case when $c$ is a sum of horizontally one-periodic and vertically one-periodic configurations, as will be shown later. We postpone the proof of \autoref{lem:rectilinear-fat-blocks} to the next section. Assuming the lemma is valid, we can finally give a proof of our main theorem.

\envparam[The main result]{theorem}{
    \label{thm-main-result}
    Let $c$ be a two-dimensional non-periodic configuration. Then
        $P_c(M,N) > MN$
    holds for all but finitely many choices $M,N \in \N$.
}

\env{proof}{
    By the discussion preceding \autoref{def:counterexample}, it is enough to consider counterexample candidates $c$. Note that either at least one of $\phi_i$ is neither horizontal nor vertical, or $\opc(c) = 2$ and the directions of $\phi_1, \phi_2$ are horizontal and vertical in some order. In either case, by \autoref{lem:fat-blocks} or \autoref{lem:rectilinear-fat-blocks}, there are $M_0, N_0$ such that for $M > M_0, N >  N_0$ we have $P_c(M,N) > MN$.
    
    Let $(m_\phi,n_\phi) = \bbox(\phi)$ and assume $n_\phi < N \leq N_0$. By \autoref{lem:thin-blocks} for each such $N$ all but finitely many $M$ satisfy $P_c(M,N) > MN$. Therefore for the whole range $n_\phi < N \leq N_0$ the condition can be violated only finitely many times. The situation for $m_\phi < M \leq M_0$ is symmetric.
    
    Finally, if $M \leq m_\phi$ or $N \leq n_\phi$ the complexity is greater than $MN$ by \autoref{lem:very-thin-blocks}. This concludes the proof.
\qed}

\env{corollary}{
    If $c$ is a two-dimensional configuration such that $P_c(M,N) \leq MN$
    holds for infinitely many pairs $M,N \in \N$, then $c$ is periodic.
}

\section{The Rectilinear Case}
\label{sec-rectilinear-case}


To complete the proof of our main result it remains to prove \autoref{lem:rectilinear-fat-blocks}. Let us restate the lemma first. Define a \emph{rectilinear} configuration to be a two-dimensional configuration which can be written as a sum of horizontally and vertically periodic configuration.

\env{lemma}{
    \label{lem:rectilinear-nonperiodic}
    Let $c$ be a finitary integral two-dimensional configuration. The following are equivalent:
    \envparam[(i)]{enumerate}{
        \item $c$ is rectilinear and non-periodic
        \item $c$ is a sum of horizontally one-periodic finitary configuration and vertically one-periodic finitary configuration
        \item $\opc(c) = 2$ and the directions of $\phi_1$ and $\phi_2$ are horizontal and vertical, in some order.
    }
}
\env{proof}{
    We prove \textit{(i)} $\Rightarrow$ \textit{(iii)} $\Rightarrow$ \textit{(ii)} $\Rightarrow$ \textit{(i)}. Assume \textit{(i)}. Since $c$ is non-periodic $\opc(c) \geq 2$. Let $m$, $n$ be the respective periods of the horizontal and vertical component of $c$. Then $c$ is annihilated by $(x^m-1)(y^n-1)$. The $\phi_i$ components are line polynomials in distinct directions dividing this polynomial. Therefore $\opc(c) = 2$, one $\phi_i$ is horizontal and the other one vertical.

    The implication \textit{(iii)} $\Rightarrow$ \textit{(ii)} follows directly from \autoref{explicit-2d-decomposition}.

    For the remaining implication assume \textit{(ii)}. Then $c$ is rectilinear, and it is also non-periodic by \autoref{ex:sum-of-oneperiodic}.
\qed}

With this notation we can restate \autoref{lem:rectilinear-fat-blocks}:

\env{lemma}{
    \label{lem:35-restated}
    Let $c$ be a finitary integral rectilinear non-periodic configuration. Then there exist positive integers $M_0,N_0$ such that for $M > M_0$ and $N > N_0$ holds $P_c(M,N) > MN$.
}


Let us give an overview of the proof. First we show that it is enough to consider \emph{binary} configurations, i.e. configurations with coefficients from $\{0,1\}$. Then, with the help of symbolic dynamics, we show that either the configuration already has a high complexity, or it contains arbitrarily large doubly periodic region. This reduces to study of configurations which are non-periodic, but vertically periodic on the upper half plane $\makeset{(x,y) \in \Z^2}{x \geq 0}$ and horizontally periodic on the right half plane $\makeset{(x,y) \in \Z^2}{y \geq 0}$. We, finally, settle this case combinatorially.

\medskip

\env{lemma}{
    \label{lem:binary}
	Let $c$ be a non-periodic configuration. Then the coefficients of $c$ can be mapped to $\{0,1\}$ such that the resulting configuration is non-periodic.
}
\env{proof}{
	First let us map a given coefficient $\alpha$ to $1$ and the rest to $0$. If any of these configurations is non-periodic we are done. Assume each of them is periodic. Because $c$ is non-periodic there must be two coefficients $\alpha, \beta$ such that the corresponding configurations are one-periodic in distinct directions. Denote their vectors of periodicity $\u, \v$ respectively.

	Observe that no sublattice modulo $\gen{\u,\v}$ in $c$ can contain both coefficients $\alpha$ and $\beta$ -- if there is $\alpha$, the whole line in direction $\u$ contains coefficients $\alpha$ and similarly for $\beta$ and a line in direction $\v$. These lines intersect, which is a contradiction.

	Define $c'$ by mapping both $\alpha$ and $\beta$ to $1$ and the rest to $0$. We will show that $c'$ is non-periodic. For contradiction suppose there is a vector of periodicity $\vec w$, by scaling it we can assume that $\vec w \in \gen{\u,\v}$.

	Now the direction of $\vec w$ differs from $\u$ or $\v$, without loss of generality assume it is different from $\u$. Then in $c$ the coefficients $\alpha$ are periodic with the vector $\vec w$ -- any line in direction $\vec w$ which contains $\alpha$ must contain only $\alpha$ and $\beta$ from the periodicity of $c'$, and the whole line lies in a sublattice modulo $\gen{\u,\v}$ and therefore contains $\alpha$ only. But the coefficients $\alpha$ are periodic also in the direction $\v$, which is in contradiction with one-periodicity.
\qed}

\env{note*}{We gave an elementary proof since the claim is not related to the theory developed in this paper. With it, however, it can be shortened. If there are two coefficients which are one-periodic in distinct directions, then the configuration obtained by mapping them to 1 and the rest to 0 is a sum of two one-periodic configurations having distinct directions. Such a configuration is, by \autoref{ex:sum-of-oneperiodic}, non-periodic.}

It is clear that by mapping the coefficients of $c$ into a configuration $c'$ the complexity can only decrease or not change: $P_c(M,N) \geq P_{c'}(M,N)$. Therefore we can restrict our efforts only to binary configuration:

\env{corollary}{If Nivat's conjecture holds for binary configurations, then it holds in general. Similarly, if \autoref{lem:35-restated} holds for binary configurations, then it holds in general.
}
\env{proof}{
    By \autoref{lem:binary} a non-periodic configuration $c$ can be mapped to a binary non-periodic configuration $c'$. Nivat's conjecture and \autoref{lem:35-restated} give a lower bound on the complexity of $c'$. The same bound applies also for $c$.
\qed}

Let us say that two configurations are \emph{disjoint} if they do not both have a non-zero coefficient at the same position, i.e. if $\supp(c_1) \cap \supp(c_2) = \emptyset$ where $\supp(c)$ is defined by $\supp(c) = \makeset{\v \in \Z^d}{c_\v \ne 0}$.

\env{lemma}{
	Let $c$ be a binary two-dimensional configuration annihilated by $(x^m-1)(y^n-1)$ for some $m,n \in \N$. Then there exist disjoint binary two-dimensional configurations $c_1, c_2$ such that $c_1$ has horizontal period $m$, $c_2$ has vertical period $n$, and $c = c_1 + c_2$.
}
\env{proof}{
    Let $\u = (m,0)$, $\v = (0,n)$. The configuration $c$ decomposes into finitely many sublattices modulo $\gen{\u,\v}$. We show that each of these sublattices is $\u$- or $\v$-periodic. The proof is then finished by setting $c_1$ to contain all the $\u$-periodic sublattices and $c_2$ to contain the rest which is necessarily $\v$-periodic.

    Let $c'$ be a binary configuration defined by $c'_{i,j} = c_{\vec{w} + i\u + j\v}$ for some $\vec{w} \in \Z^2$, i.e. $c'$ is one of the sublattices "condensed."	Then $c'$ is annihilated by $(x-1)(y-1)$. The only possible $2 \times 2$ blocks in such a configuration are
	\env{align*}{
    	\env{bmatrix}{
    	    0 & 0 \\
    	    0 & 0
    	},
    	\env{bmatrix}{
    	    0 & 0 \\
    	    1 & 1
    	},
    	\env{bmatrix}{
    	    1 & 1 \\
    	    0 & 0
    	},
    	\env{bmatrix}{
    	    0 & 1 \\
    	    0 & 1
    	},
    	\env{bmatrix}{
    	    1 & 0 \\
    	    1 & 0
    	},
    	\env{bmatrix}{
    	    1 & 1 \\
    	    1 & 1
    	}.
	}
	Notice that if there is $0$ adjacent to $1$ in one row then there is an all-zero and all-one column. Similarly, if a column contains adjacent $0$ and $1$ then there is an all-zero and all-one row. These two options cannot happen simultaneously. Therefore $c'$ is $(1,0)$- or $(0,1)$-periodic, which means that the corresponding sublattice in $c$ is $\u$- or $\v$-periodic.
\qed}

To proceed we need some basic concepts of symbolic dynamics, for a comprehensive reference see \cite{Kurka}. The {\em orbit closure\/} $\overline{{\cal O}(c)}$ of a configuration $c$ is the subshift it generates:
$\overline{{\cal O}(c)}$ contains precisely those configurations $c'$ whose finite patterns are among the finite patterns of $c$. If $\overline{{\cal O}(c')}=\overline{{\cal O}(c)}$ for all $c' \in \overline{{\cal O}(c)}$ then the subshift is minimal. This happens if and only if $c$ is {\em uniformly recurrent}, that is, if and only if for every finite pattern $p\in A^D$ that appears somewhere in $c$ there exists finite $E\subseteq \Z^d$ such that every $E$-pattern of $c$ contains $p$ as a  subpattern.

Note that for all $c' \in \overline{{\cal O}(c)}$ and all finite $D\subseteq \Z^d$ we have $P_{c'}(D)\leq P_{c}(D)$.
We can replace configuration $c$ in our proof with any non-periodic $c'$ from
its orbit closure without increasing the complexity.

We use the following one-dimensional technical lemma.
\env{lemma}{
    \label{lem:onedimlemma}
	Let $c\in A^{\Z}$  be a non-periodic one-dimensional configuration, and let $X=\overline{{\cal O}(c)}$ be its orbit closure. Then one of the following holds:
	\begin{itemize}
    	\item[(a)] $X$ contains a uniformly recurrent element that is not periodic, or
	    \item[(b)] $X$ contains some $c_R$ that is non-periodic but is eventually periodic on the right.
	\end{itemize}
}

\env{proof}{
Suppose that (a) does not hold. Let us prove the existence of $c_R$.

Consider the sequence $c, \sigma(c), \sigma^2(c), \dots$ of configurations, where $\sigma$ is the left shift. The sequence has an accumulation point $c'$ under the standard compact topology of $A^\Z$. Then $c'$ is in $X$ and, in fact, every finite pattern that appears in $c'$ appears arbitrarily far on the right in $c$.

It is well known that every subshift contains a uniformly recurrent configuration. Let $c''$ be a uniformly recurrent configuration in the orbit closure $\overline{{\cal O}(c')}$ of $c'$. Then $c''$ is also in $X$ and, in fact, every finite word that appears in $c''$ appears in $c'$ and hence appears arbitrarily far on the right in $c$.

Because (a) does not hold, $c''$ is periodic. It is annihilated by polynomial
$f(x)=(x^n-1)$ for some $n\geq 1$. This means that $fc$ contains arbitrarily long segments of $0$'s arbitrarily far on the right. Because $fc\neq 0$, the segments of $0$'s sufficiently far on the right have period breaks: non-zero values followed by arbitrarily long runs of $0$'s. We obtain $c_R$ by translating $c$ in such a way that the period breaking points are at the position $-1$ and take an accumulation point of these translates for longer and longer runs of $0$'s. We have that $fc_R$ is zero at all non-negative positions, but non-zero at $-1$.
\qed}

The next lemma is a two-dimensional variant of the lemma above. It allows us to replace $c$ by a more convenient configuration from its orbit closure. Recall the notation $c(\v+D)$ which is the same as $c_{\v+D}$, that is, the pattern $D$ extracted from position $\v$ in $c$. Let us denote $[n] := \{0, 1, \dots, n-1\}$, then we can concisely write $[m]\times[n]$ for the $m \times n$ block.

\env{lemma}{
    \label{lem:twodimlemma}
    Let $\vec{u_1}$ be horizontal, $\vec{u_2}$ vertical vector and let $c$ a non-periodic binary configuration which can be written as a disjoint sum $c_1 + c_2$ where $c_i$ has period $\vec{u_i}$. Then there is a non-periodic $c' \in \overline{{\cal O}(c)}$ which can be written as a disjoint sum $c'_1 + c'_2$ where $c'_i$ has period $\vec{u_i}$, and one of the following two possibilities holds:
    \begin{itemize}
    \item[(a)] $c'_1$ or $c'_2$ is uniformly recurrent, or
    \item[(b)]
$c'_1$ is doubly periodic on the upper half plane $\{(x,y)\in\Z^2\ |\ y \geq 0\}$, and\\
$c'_2$ is doubly periodic on the right half plane $\{(x,y)\in\Z^2\ |\ x \geq 0\}$.
    \end{itemize}
}
\env{proof}{
    Denote $\vec{u_1} = (m,0)$ and $\vec{u_2} = (0,n)$, $m,n > 0$. The idea is to partition $\Z^2$ into $m\times n$ blocks, consider $c_1$ and $c_2$ as one-dimensional configurations over such blocks, and apply \autoref{lem:onedimlemma}.

    Let $D = [m]\times[n]$ be the $m \times n$ block and let $A=\{0,1\}^D$. Construct the following one-dimensional configurations $e_1$ and $e_2$ over alphabet $A$: For all $k \in \Z$,
\begin{equation}
	\label{eq:2d_to_1d}
	\begin{array}{rcl}
    	e_1(k) &=& c_1(k\vec{u_2}+D),\\
	    e_2(k) &=& c_2(k\vec{u_1}+D).
	\end{array}
\end{equation}
The sequence $e_1$ encodes a vertical stripe of width $m$ in $c_1$ which, by $\vec{u_1}$-periodicity, determines $c_1$. Similarly, $e_2$ encodes a horizontal stripe of height $n$ in $c_2$ which determines $c_2$.

Because $c_i$ is not doubly periodic, the configuration $e_i$ is non-periodic. We can apply \autoref{lem:onedimlemma} on $e_i$ to obtain $e'_i\in \overline{{\cal O}(e_i)}$. Let us reconstruct a two-dimensional configuration $c'_i$ from $e'_i$ by the inverse of (\ref{eq:2d_to_1d}): Let $c'_i$ be $\vec{u_i}$-periodic and for all $k \in \Z$,
\env{align*}{
	c'_1(k\vec{u_2}+D) &= e'_1(k),\\
	c'_2(k\vec{u_1}+D) &= e'_2(k).
}
Because $e'_i$ is non-periodic, $c'_i$ is one-periodic. From $e'_i\in \overline{{\cal O}(e_i)}$ follows $c'_i \in \overline{{\cal O}(c_i)}$. More precisely, for any $E \subset \Z^2$ exist $k_1, k_2 \in \Z$ such that
\env{align*}{
    c'_1(E) &= c_1(k_2 \vec{u_2} + E),\\
    c'_2(E) &= c_2(k_1 \vec{u_1} + E).
}
Restricting $E$ to $\{\v\}$ gives that $c'_1$ and $c'_2$ are disjoint since $c'_i(\v) = c_i(k_1 \vec{u_1} + k_2 \vec{u_2} + \v)$.

Now set $c' = c'_1 + c'_2$, by \autoref{lem:rectilinear-nonperiodic} it is a non-periodic configuration. We claim that $c' \in \overline{{\cal O}(c)}$: Indeed, if $E \subset \Z^2$ is arbitrary, then there exist $k_1, k_2 \in \Z$ such that
\env{align*}{
    c'(E) &= c'_1(E) + c'_2(E) = c_1(k_2 \vec{u_2} + E) + c_2(k_1 \vec{u_1} + E) \\
          &= c_1(k_1 \vec{u_1} + k_2 \vec{u_2} + E) + c_2(k_1 \vec{u_1} + k_2 \vec{u_2} + E) \\
          &= c(k_1 \vec{u_1} + k_2 \vec{u_2} + E).
}

It remains to prove that one of the cases (a) or (b) holds. If one of $e'_i$ is uniformly recurrent, so is $c'_i$, and the case (a) holds. Otherwise, by \autoref{lem:onedimlemma}, $e'_1$ and $e'_2$ are eventually periodic to the right, which implies that the corresponding two-dimensional configurations $c'_1$ and $c'_2$ are doubly periodic on the upper half plane and on the right half plane, respectively. In that case (b) holds.
\qed}

\env{lemma}{
	Let $c' = c_1' + c_2'$ as in \autoref{lem:twodimlemma}. If $c_1'$ or $c_2'$ is uniformly recurrent, then for $M,N$ large enough we have
	\env{align*}{
		P_{c'}(M,N) > MN.
	}
}
\env{proof}{
	Without loss of generality assume $c_1'$ is uniformly recurrent. Consider a sublattice $\Lambda$ modulo $\gen{\vec{u_1},\vec{u_2}}$ in $c'$. Because $c'$ is a disjoint sum and the vectors of periodicity of $c_1',c_2'$ are $\vec{u_1},\vec{u_2}$ respectively, $c'$ restricted to $\Lambda$ is identical with one of $c_1'$ or $c_2'$.

	Let us assume that $c_1'$ is not constant 1 on any sublattice $\Lambda$ -- if it is, we can subtract this sublattice from $c_1'$ and add it to $c_2'$. Note that this does not change uniform recurrence of either configuration. 
	
	Since $c_1'$ is uniformly recurrent, it is also uniformly recurrent when restricted to any sublattice $\Lambda$. Let $M,N$ be large enough such that every block $M \times N$ in $c_1'$ contains at least one 0 and one 1 from each sublattice of $c_1'$ on which it is not constant zero.

	Let $D = [M]\times[N]$ and consider a block pattern $c'_{\v+D}$. Let $\Lambda$ be a sublattice. We know that $c'_{\v+D}$ agrees with $c'_{1,\v+D}$ or $c'_{2,\v+D}$ on $\Lambda$. We claim that the former happens if and only if $c'_{\v+D}$ restricted to $\Lambda$ is either constant zero or if it is not $\vec{u_2}$-periodic -- this is because $M,N$ were chosen such that on sublattices which contain 1, no restriction of $c'_{1,\v+D}$ to $\Lambda$ is $\vec{u_2}$-periodic, while all restrictions of $c'_{2,\v+D}$ are.

	In other words, from a block $c'_{\v+D}$ we can determine the blocks $c'_{1,\v+D}$ and $c'_{2,\v+D}$. By \autoref{cor:disjoint-columns}, if $N$ is large enough there are $M$ disjoint columns of blocks $M \times N$ in $c_2'$. Because $c_1'$ is vertically non-periodic, by Morse-Hedlund theorem each column of blocks $M \times N$ in $c_1'$ contains at least $N+1$ distinct blocks. By positioning the block in $c'$ these can be combined to achieve the lower bound
	\env{align*}{
		P_{c'}(M,N) \geq M(N+1) > MN.
	}
\qed}

The remaining case to study is illustrated in \autoref{fig:quadrant}.

\envparam{figure}{
    \centering
    \includegraphics[scale=0.65]{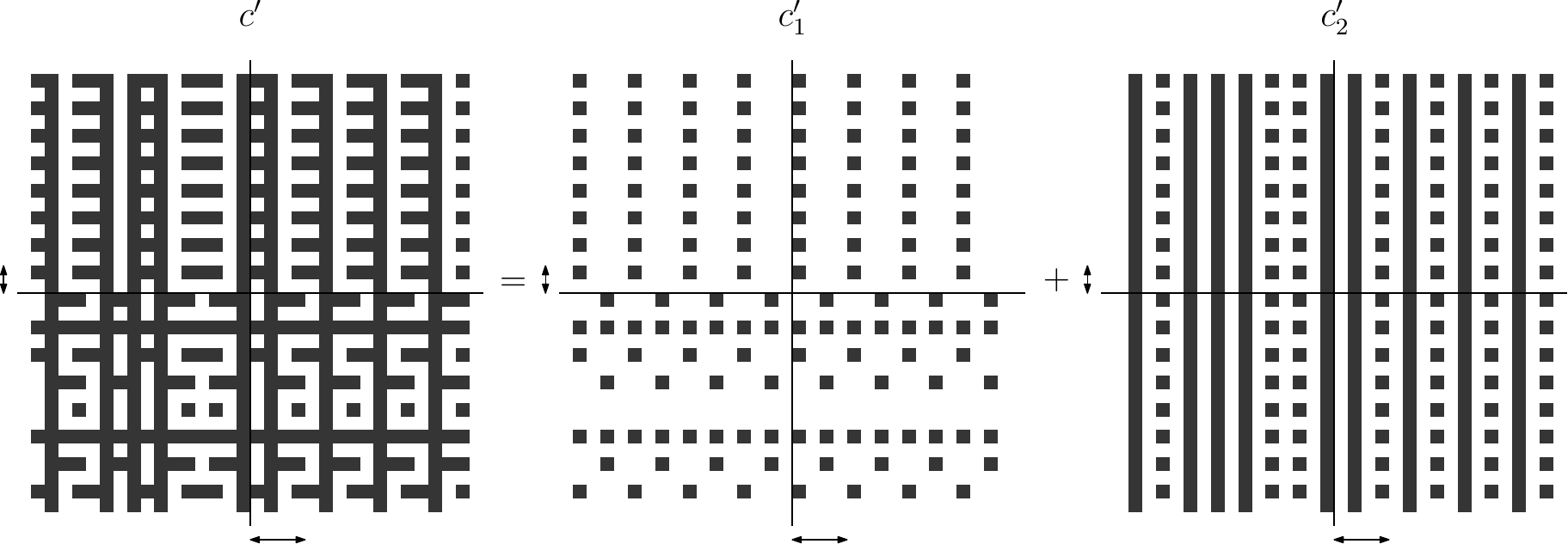}
    \caption{The binary configuration $c'$ is nonperiodic, but periodic horizontally on the right half plane with period 4 and periodic vertically on the upper half plane with period 2. It can be decomposed into a disjoint sum of two periodic configurations $c'_1 + c'_2$ which are one-periodic, but doubly periodic on the upper and right half plane, respectively. (The color black corresponds to coefficient 1 and white to 0.)}
    \label{fig:quadrant}
}

\env{lemma}{
	Let $c' = c_1' + c_2'$ as in \autoref{lem:twodimlemma} and the case (b) holds. Then for $M,N$ large enough we have
	\env{align*}{
		P_{c'}(M,N) > MN.
	}
}
\env{proof}{
    The configuration $c'$ is vertically periodic on the upper half plane, let $n_0 \in \N$ be the shortest vertical period. Let us call a point $\v \in \Z^2$ \emph{period-breaking} if $c'_{\v} \ne c'_{\v+(0,n_0)}$. Let $\vec{v_0} = (x_0, y_0)$ be a topmost period-breaking point (i.e. with maximal $y_0$).
    
    We know that $c'$ is annihilated by $(x^m-1)(y^n-1)$ for some $m,n \in \N$. We claim that every point $\vec{v_0}+(km,0)$ for $k \in \Z$ is period-breaking. By the choice of $\vec{v_0}$ we have $c'_{\vec{v_0}} \ne c'_{\vec{v_0}+(0,n_0)} = c'_{\vec{v_0}+(0,n_0n)}$. In particular, $(y^{n_0n}-1)c' \ne 0$ because it has a non-zero value at $\vec{v_0}$. Note that $(x^m-1)$ annihilates $(y^{n_0n}-1)c'$, so this configuration has a horizontal period $m$ and therefore for all $k \in \Z$
    $$
        c'_{\vec{v_0}+(km,0)} \ne c'_{\vec{v_0}+(km,n_0n)} = c'_{\vec{v_0}+(km,n_0)},
    $$
    as claimed.
    
    Let $M > m, N > \max\{n_0, n\}$. Consider a row of blocks $M \times N$ which overlaps the rows $y_0$ and $y_0+n_0$, there are $N-n_0$ such rows. Any block in these rows sees a period-breaking point, and we can distinguish between the blocks in distinct rows by the highest row inside the block where the period-breaking occurs. Therefore there are $N-n_0$ disjoint lines of blocks in the direction $(1,0)$. Using \autoref{lem:complex-lines}(b) with the roles of horizontal and vertical exchanged, each of these lines contains at least $M+1$ distinct blocks which gives altogether
    $$
        (M+1)(N-n_0)
    $$
    distinct blocks.
    
    We will find additional blocks inside the half plane $U := \makeset{(x,y) \in \Z^2}{y > y_0}$, all such blocks are distinct from those already counted since there is no period-breaking point inside them. Configuration $c'$ restricted to $U$ is not periodic horizontally, but it is periodic horizontally on the right half plane with minimal period $m_0 \in \N$. Define a point $\v \in \Z^2$ \emph{horizontal period-breaking} if $c'_{\v} \ne c'_{\v+(m_0,0)}$ and let $\v_1 = (x_1, y_1)$ be a rightmost such point. Clearly, $\v + (0, kn_0)$ is a horizontal period-breaking point for all $k \in \Z$.
    
    Because the minimal vertical period of $c'\!\!\upharpoonright_U$ is $n_0$, there is a finite set of columns $\{x_2, \dots, x_t\}$ such that their joint vertical period is $n_0$. Let $S = \{x_1, x_1+m_0, x_2, \dots, x_t\}$ and set $m' = \max S - \min S$.
    
    If $M > m'$, then a block $M \times N$ can be positioned such that it overlaps with all the columns in $S$. There are $M-m'$ distinct horizontal positions when this happens, and they can be identified by the rightmost horizontal period-breaking point. For each of them we can slide the block up into $n_0$ positions with distinct patterns, giving $(M-m')n_0$ distinct blocks. Altogether we have
    $$
        P_{c'}(M,N) \geq (M+1)(N-n_0) + (M-m')n_0 = MN + N - m' n_0.
    $$
    It suffices to choose $M > \max\{m,m'\}$ and $N > \max\{n,n_0,m' n_0\}$ to finish the proof.
\qed}

The proof of \autoref{lem:rectilinear-fat-blocks} follows by putting together all the lemmas in this section. That completes the proof of our main result, \autoref{thm-main-result}.


\printbibliography[notcategory=excluded]

\end{document}